\documentclass{article} 
\usepackage{iclr2027_conference,times}

\usepackage{amsmath,amsfonts,bm}

\def\eqref#1{equation~\ref{#1}}

\def\1{\bm{1}}

\DeclareMathAlphabet{\mathsfit}{\encodingdefault}{\sfdefault}{m}{sl}
\SetMathAlphabet{\mathsfit}{bold}{\encodingdefault}{\sfdefault}{bx}{n}

\usepackage{hyperref}
\usepackage{url}

\title{Variational Trajectory Flow Matching with a Non-Linear Interpolant}

\author{Gurjeet Sangra Singh$^{1,2}$, Frantzeska Lavda$^{2}$, Alexandros Kalousis$^{2}$  \\ 
$^{1}$ University of Geneva, Department of Computer Science \\
$^{2}$ University of Applied Sciences Western Switzerland, HES-SO Geneva \\
\texttt{gurjeet.singh@etu.unige.ch},\\
\texttt{\{frantzeska.lavda,alexandros.kalousis\}@hesge.ch} \\
}

\usepackage{amstext}
\usepackage{amsthm}

\usepackage{amsmath}
\usepackage{xcolor}
\usepackage{comment}
\usepackage{tikz}
\usetikzlibrary{decorations.pathmorphing}
\usepackage{multirow}
\usepackage{booktabs} 
\usepackage{caption} 
\usepackage{graphicx} 
\usepackage[linesnumbered,ruled,vlined]{algorithm2e}

\usepackage{enumerate}

\usepackage{stfloats}
\usepackage{subcaption} 
\newcommand{\vect}[1]{\boldsymbol{#1}}

\newcommand{\expnumber}[2]{{#1}\mathrm{e}{#2}}
\DeclareMathOperator{\logU}{log-\mathcal{U}}

\definecolor{darkgreen}{HTML}{2E7D32}
\definecolor{salmon}{rgb}{0.98,0.50,0.45}
\definecolor{grey}{gray}{0.6}

\iclrfinalcopy 
\begin{document}

\maketitle
\lhead{Preprint} 

\begin{abstract}
Simulation-free methods such as Trajectory Flow Matching learn dynamics by regressing the time derivative of a local interpolant placed between consecutive observations. The interpolant defines the regression target, and under sparse, irregular sampling, it underrepresents the dynamics and bounds what can be learned. We show that no fixed interpolant dominates: the best of linear, quadratic, spline and quintic changes with the system and with the sampling density. We propose to learn the interpolant instead. An amortized implicit neural representation (INR) is fitted to the training trajectories and frozen as a teacher supplying the target velocity to a learnable velocity field. We further frame the velocity field within a variational framework, yielding a probabilistic forecaster that, at inference, requires only the velocity field and no SDE solver. We evaluate our method on a range of dynamical systems: it consistently achieves high accuracy and more stable long-horizon rollouts under sparser observations, at matched velocity-field capacity.
\end{abstract}

\section{Introduction}

Modelling dynamical systems is challenging across scientific domains, particularly when states evolve in high-dimensional spaces, as in fluid dynamics, climate science, biology, neuroscience, and engineering.
Flow Matching-based methods \citep{lipman2023,albergo2023,liu2023flow,tong2024improving} are continuous-time generative models with the ability to capture complex data distributions. They bypass the need for a numerical solver during training and avoid differentiating through it, which improves memory scalability and makes high-dimensional settings more tractable, whereas simulation-based approaches \citep{chen2018neuralode, kidger2020, dupont2019ode} can be more constrained. Recent simulation-free extensions include Trajectory Flow Matching \citep[TFM]{zhang2024tjfm}, proposed for dynamics learning and time-series forecasting.

Unlike standard Flow Matching, TFM learns \emph{state-to-state} dynamics rather than a noise-to-state or noise-to-trajectory transformation \citep{fu2025moflowonestepflowmatching,huang2026recursiveflowmatching}. State-to-state dynamics preserves temporal causality: the velocity at a query time describes the physics of the real system, not an unrelated velocity artifact induced by a Gaussian probability path from noise to data. This makes TFM a natural framework for non-linear, high-dimensional continuous-time dynamics.

However, real-world observations are often \emph{sparse} and \emph{irregular} in time: a sensor collects measurements at various time points and under diverse conditions when possible, and numerical solvers are run at different step sizes for compute cost. In this regime, trajectory-matching methods rely on an often implicit modelling decision: they construct a probability path between observations and regress its analytical time derivative. The chosen interpolant is not merely a numerical detail: it determines the target velocity used for training and constrains which dynamics can be represented. In particular, the linear interpolant used in TFM provides a first-order approximation of the trajectory between consecutive observations. It is accurate only when the interval is short relative to the curvature of the dynamics. Such an assumption is often violated by real systems that are curved and their velocity may change between samples. The resulting velocity of TFM is then constant over the interval, whereas the true velocity is not.

\begin{figure}[h!]
    \centering
    \begin{tabular}{@{}ll@{}}
        \vtop{\vskip-10.0ex\hbox{\textbf{(a)}}} & 
        \includegraphics[width=0.86\linewidth]{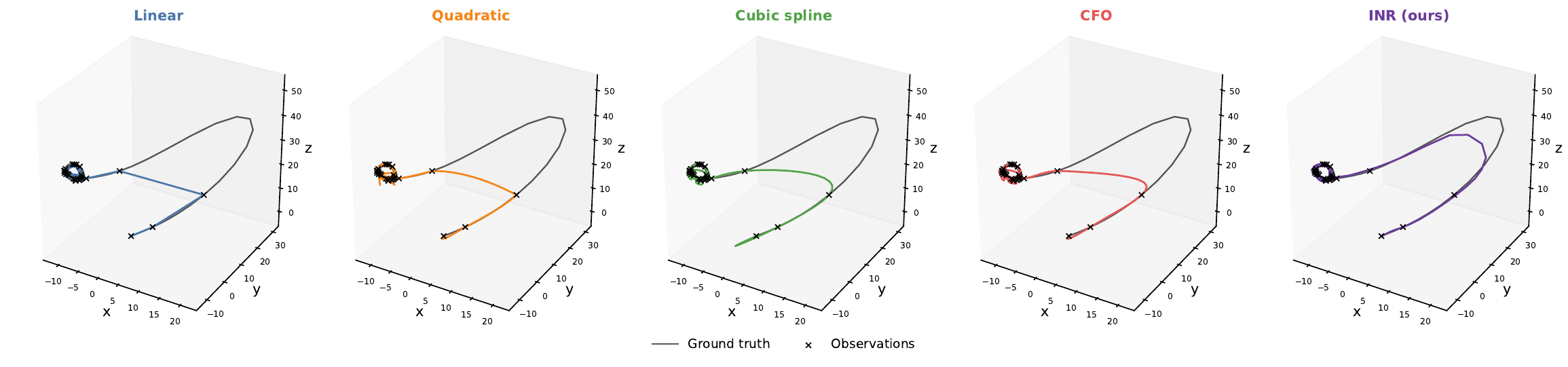} \\ 
        
        \vtop{\vskip-18.0ex\hbox{\textbf{(b)}}} & 
        \includegraphics[width=0.86\linewidth]{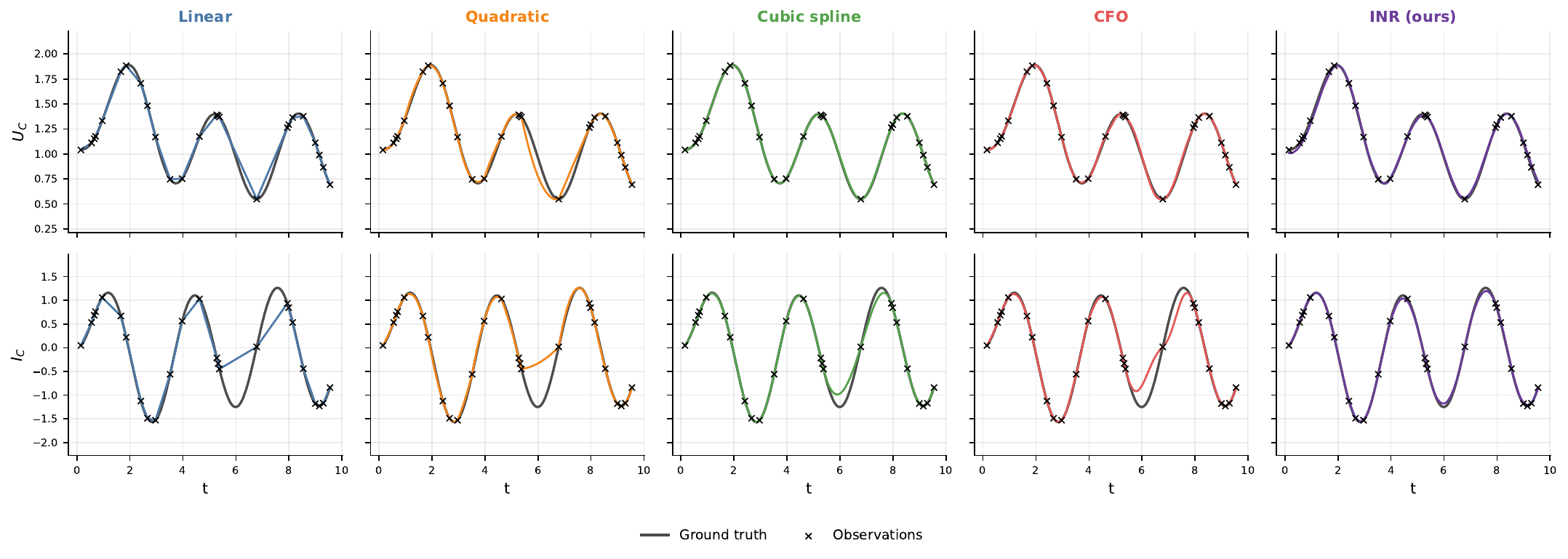}
    \end{tabular}
    \caption{Effect of the interpolant choice at $25\%$ observation density on Lorenz (a) and RLC (b) test sample trajectories. Crosses (x) in the figures are the
    observations every method receives. A linear interpolant draws straight-line chords across large gaps; other fixed interpolants do better, but still fit the data poorly. Our t-INR approach learns both global and local structure and tracks the trajectory under a sparse regime.}
    \label{fig:interp_fail_ex}
\end{figure}

This limitation has motivated several variants of TFM for sparse or irregularly sampled data. In this work, we review these methods within a unified framework and investigate how their interpolant assumptions affect the learned dynamics. TFM relies on linear interpolation; later methods replace it with quadratic \citep[LFM]{islam2026longitudinal}, cubic spline \citep[MMFM]{rohbeck2025,lee2025multimarginal}, or quintic spline \citep[CFO]{hou2025cfo}. These methods can provide more flexible local approximations, but they keep the same underlying assumption: the interpolant is fixed in advance and fitted locally from only a few neighbouring observations. As a result, the learning targets vary with the sampling times, and this dependence can be significant when observations are sparse. Figure~\ref{fig:interp_fail_ex} illustrates this failure mode.

A complementary line of work incorporates known physical expressions. Physics-informed generative models impose differential-equation constraints as soft regularizers \citep{sholokhov2023,huang2024diffusionpde,baldan2026} for learning forward models, while hybrid models embed known dynamics into the model architecture for forecasting tasks \citep{naoya_2021,qian2021pharma_ode,yin_2021,naoya_2023,wehenkel2023robust,singh2026variational}. These approaches operate in a distinct setting: they assume access to, or exploit, explicit physical information. Other relevant work focuses on large-scale PDEs with varying geometries using spatial mesh-free methods, such as Neural Operators \citep{deeponet2021,fno2021,neuralfield2022,serrano2023operator,gino2023}, which learn mappings between functional spaces rather than on discrete fixed grids. In this paper, we study the limitations introduced by fixed interpolation targets in simulation-free methods and propose a solution for learning dynamics from irregularly sampled trajectories.

\paragraph{Our Contributions.}
\begin{itemize}
    \item We identify fixed interpolation as the core assumption in simulation-free trajectory matching under sparse, irregular sampling, and show experimentally that no fixed interpolant dominates across systems or sampling regimes.
    
    \item We propose to learn the interpolant by an amortized t-INR that imposes no
    assumption on the spacing between observations.  We further show how the learned t-INR can be used as a frozen teacher to provide target velocity.
    \item We derive a variational velocity field trained from these teacher targets, giving a probabilistic forecaster without an SDE formulation or an SDE solver at inference.
    \item Across five dynamical systems and one real-world experiment, we show that our method improves the fixed-interpolant baselines, with the largest gains in the sparse regime.
\end{itemize}

\section{Preliminaries} \label{sec:background}
\subsection{Problem setting} \label{sec:problem_settings}
We are given a collection of trajectories $\mathcal{D}=\{\mathbf{X}^{(i)}\}_{i=1}^N$. Each sample $\mathbf{X}^{(i)}$ corresponds to a sequence of observed states $\mathbf{X}^{(i)} = \left(\mathbf{x}^{(i)}_{t_0}, \dots, \mathbf{x}^{(i)}_{t_T}\right)$, where $\mathbf{x}^{(i)}_{t_{k}} \in \mathcal{X} \subseteq \mathbb{R}^d$ denotes the state at each discretized time instant $t_{k}$. We also denote a segment of a trajectory as $\mathbf{X}_{t_{k-h:k+1}}=\left(\mathbf{x}_{t_{k-h}}, \dots, \mathbf{x}_{t_{k}}, \mathbf{x}_{t_{k+1}} \right)$, i.e., a sequence of consecutive points. To reproduce some real scenario case of measurements, we assume points of the trajectory are measured at sparse random (non-equidistant) time points $t_0 < t_1 < ... < t_T$ where $t_{k} \in \mathbb{R}$, and $T \in \mathbb{N}$.
These trajectories are assumed to be independent realisations from an unknown \textit{data} distribution $\pi(\mathbf{X}) \in \mathcal{P}(\mathcal{X}^{T+1})$, where $\mathcal{P}(\mathcal{X}^{T+1})$ is the set of all probability measures over the trajectory space $\mathcal{X}^{T+1}$. The objective is to learn the dynamical system for forecasting tasks in which, given a current state $\mathbf{x}_{t_k}$ and an arbitrary history of the past sparse observations  $\mathbf{X}_{t_{k-h:k}}$, generate plausible future trajectories that approximate the true conditional distribution $\pi(\mathbf{X}_{>t_{k}} \mid \mathbf{X}_{t_{k-h:k}})$ for any future time $t>k$.

\subsection{Trajectory Flow Matching}
Trajectory Flow Matching (TFM) methods ~\citep{zhang2024tjfm,islam2026longitudinal,rohbeck2025}, extend training simulation-free methods like Conditional Flow Matching \citep[CFM]{lipman2023} and variants \citep{albergo2023,liu2023flow} to dynamical systems where data come as time-series trajectories $\mathbf{X} = (\mathbf{x}_{t_0}, \mathbf{x}_{t_1}, \ldots, \mathbf{x}_{t_T})$. The modelled dynamics is the following Stochastic Differential Equation (SDE):
\begin{equation}
    d\mathbf{x}_{t} = v(\mathbf{x}_t, t) \; dt + g(\mathbf{x}_t, t) \; dW_t, 
    \label{eq:ito-dynamics}
\end{equation}

where $v: \mathcal{X} \times [0,T] \mapsto \mathbb{R}^d$ is the drift term, $g: \mathcal{X} \times [0,T] \mapsto \mathbb{R}^{d^2}$ is the diffusion coefficient, and $W_t$ is the $d$-dimensional Wiener-Brownian process. Under such SDE formulation, when evolving an initial density $p_0(\mathbf{x}_0)$ produces a collection of marginal density $p_t(\mathbf{x_t})$ for $t \in [0, T]$, induced by the Fokker–Planck equation:
\begin{equation}
    \partial_t p_t = - \nabla \cdot (p_t \; v_t) + \frac{g^{2}}{2} \Delta p_t.
\end{equation}

Rather than mapping from noise-to-data as CFM (where the source distribution $p_0$ is modelled as a Standard Gaussian distribution), TFM learns state-to-state dynamics, where the velocity field describes the system’s temporal evolution. To learn the dynamics in a training simulation-free approach, they first construct a Gaussian probability path between consecutive trajectory states $\mathbf{x}_{t_k}$ and $\mathbf{x}_{t_{k+1}}$ at observed time points $(t_k, t_{k+1})$. In the linear case, within the work of TFM \citep{zhang2024tjfm}, the probability path is constructed as:
\begin{align}
&p_\tau(\mathbf{x}_\tau \mid \mathbf{X}_{t_{k:k+1}}) = \mathcal{N}\!\left(\mu_\tau,\, \sigma_\tau^{2}\mathbf{1} \;\middle|\; \mathbf{X}_{t_{k:k+1}}\right), \label{eq:prob-path} \\
&\mu_\tau = \frac{(t_{k+1}-\tau)\,\mathbf{x}_k + (\tau - t_k)\,\mathbf{x}_{k+1}}{t_{k+1} - t_k}, \quad
u_\tau = \frac{\mathbf{x}_{k+1}-\mathbf{x}_{k}}{t_{k+1} - t_k}, \quad
\sigma_\tau = \sigma_0\,(\tau - t_{k})\,\frac{t_{k+1}-\tau}{t_{k+1} - t_k}. \nonumber
\end{align}
Therefore, learning the velocity field leads to a simple regression problem: 
\begin{align}
\mathcal{L}_{\text{TFM-v}}(\phi) &=
\mathbb{E}_{\substack{\pi(\mathbf{X}) \\ \pi(\mathbf{X}_{t_{k-h:k+1}} | \mathbf{X}) \\ \tau \sim \mathcal{U}(t_{k}, t_{k+1})}} 
 \left \| v_{\phi}(\mathbf{x}_\tau, \tau, \mathbf{X}_{t_{k-h:k}} ) - u(\tau; \mathbf{X}_{t_{k:k+1}}) \right\|^2 ,
\label{eq:det-TFM}
\end{align}
where $\mathbf{x}_{\tau} \sim p_\tau(\mathbf{x}_\tau | \mathbf{X}_{t_{k:k+1}})$ and $u(\tau; \mathbf{X}_{t_{k:k+1}})$ are from \eqref{eq:prob-path}, and $\pi(\mathbf{X}_{t_{k-h:k+1}} | \mathbf{X})$ is the empirical distribution of consecutive points, which includes the next, current, and history points.
However, the above loss yields a deterministic model. To learn the uncertainty of the dynamics, \cite{zhang2024tjfm} and its variants  methods \citep{islam2026longitudinal, rohbeck2025, lee2025multimarginal} propose to learn jointly $v_\phi$ and the diffusion term $g_\phi$, optimizing:
\begin{equation}
\resizebox{0.9\textwidth}{!}{%
$\begin{aligned}
\mathcal{L}_{\text{TFM}}(\phi) &=
\mathbb{E}_{\substack{\pi(\mathbf{X}) \\ \pi(\mathbf{X}_{t_{k-h:k+1}} | \mathbf{X}) \\ \tau \sim \mathcal{U}(t_{k}, t_{k+1})}}  \left\| g_\phi (\mathbf{x}_\tau, \tau, \mathbf{X}_{t_{k-h:k}} ) - 
 \left \| v_{\phi}(\mathbf{x}_\tau, \tau, \mathbf{X}_{t_{k-h:k}} ) - u(\tau; \mathbf{X}_{t_{k:k+1}}) \right\|^2 \right\|^2.
\end{aligned}$
}
\label{eq:TFM}
\end{equation}
Based on this formulation, all these variants \citep{islam2026longitudinal, rohbeck2025, lee2025multimarginal,hou2025cfo} differ mainly in $\mu_\tau$ with a fixed non-linear interpolation (quadratic or spline, quintic-spline). Instead, we make $\mu_{\tau}$ learnable.

\section{Method} \label{sec:method}
We formalize our method in the three main building blocks. We first addressed the fixed-interpolant
assumption by learning it with a temporal INR (\S\ref{sec:tinr}). We
then freeze it and use it as a teacher that supplies the velocity field  (\S\ref{sec:teacher}). Finally we make the velocity field variational (\S\ref{sec:varvf}), which gives a probabilistic forecaster that needs no SDE formulation. We also provide the pseudocode of training and inference methods in the App. \ref{app:algorithms}.

\subsection{Learning the Interpolant via INR} \label{sec:tinr}
To address the fixed interpolant assumption of TFM and variants, we introduce a learnable functional, capable of interpolating trajectory points under sparse, irregular time-step settings. We achieve this by amortizing a Implicit Neural Representation \citep[INR]{sitzmann2020siren} on the temporal axis, which learns a non-linear, smooth interpolation between sparse observations, by optimizing the following simple state $\mathbf{x}_{t_k}$ reconstruction loss:
\begin{equation}
    L_{\text{t-INR}}(\xi, \eta)= \mathbb{E}_{\pi(\mathbf{X})} \frac{1}{T} \sum_{k=0}^T \left\vert\left\vert \mathbf{x}_{t_{k}} - I_\xi (t_{k}; E_\eta (\mathbf{X})) \right\vert  \right\vert,
\end{equation}
where $\mathbf{X}$ is the observed sequence at discretised sparse points $t_k$ with state $\mathbf{x}_{t_k}$, and the temporal INR (t-INR) model compromises of: 
\begin{itemize}
    \item Encoder $E_{\eta}: \mathbf{X} \mapsto \mathbf{Z}$. It maps any discretized sequence points to a fixed dimensional latent representation of the trajectory $\mathbf{Z} \in \mathbb{R}^{q}$ (with $q < d$).
    \item Temporal Decoder $I_{\xi}: t \times \mathbf{Z} \mapsto \hat{{\mathbf{x}}}_t$. It maps any time query point $t \in [0, T]$ to a predicted state $\hat{{\mathbf{x}}}_t$, using the encoded representation from $E_\eta$. The decoder model is inspired by the SiREN architecture \citep{sitzmann2020siren}, where the temporal axis is mapped in high-dimensional space using the sine as a periodic activation function capturing temporally varying signals, including high-frequency details.
\end{itemize}

The t-INR is amortized through an encoder that processes the sparse training data, and smoothness is enforced by the siren-architecture. Thanks to amortization across trajectories and over time states (enabled by the encoder–decoder architecture), the model captures both \emph{local} and \emph{global} patterns in the training trajectories and can generalize better to unobserved time points along a trajectory something that local interpolants cannot achieve.

The t-INR is trained as a pre-training stage before solving the forecasting problem. It will be used to query at any time inside the training time range, so it is not restricted to the observed discrete grid. 

\subsection{The Interpolant as a Teacher} \label{sec:teacher}
After training the t-INR, we freeze it and \textbf{never} update it again, we denote the frozen parameters with $(\xi^{\star}, \eta^{\star})$ to improve clarity in the text. The t-INR plays the role of a teacher that supplies three quantities: \textbf{query state}, \textbf{target velocity}, and \textbf{history buffer}. Query times are drawn inside the observed span of each trajectory, $\tau \sim \mathcal{U}(t_0, t_T)$, so the teacher is always evaluated in interpolation. For any such $\tau$ it provides the pair of query state and target velocity $({\hat{\mathbf{x}}}_{\tau}, \hat{\mathbf{u}}_\tau)$:
\begin{equation}
    {\hat{\mathbf{x}}}_{\tau} \leftarrow I_{\xi^{\star}}(\tau; E_{\eta^{\star}}(\mathbf{X})),\quad \text{and} \quad  \hat{\mathbf{u}}_\tau \leftarrow \partial_{\tau} I_{\xi^{\star}}\big(\tau; E_{\eta^{\star}}(\mathbf{X})\big).
    \label{eq:inr-target}
\end{equation}
We compute $\hat{\mathbf{u}}_\tau$ by efficient automatic differentiation (see App.~\ref{app:autodiff} for more practical details). We remark that our target velocity is not \emph{constant} on an interval and does not depend only on local points, whereas the linear interpolant yields a constant velocity (\eqref{eq:prob-path}).

Because the teacher can be queried anywhere inside the observed span, we do not restrict conditioning to the observed grid: we place the history buffer at $h$ times before $\tau$ whose \emph{gaps} are drawn log-uniformly, and read the corresponding states from the teacher,
\begin{equation}
    \label{eq:history_buffer}
    \hat{\mathbf{H}}_{\tau} = \big(\hat{\mathbf{x}}_{s_1}, \dots, \hat{\mathbf{x}}_{s_h}\big),
    \;\;
    \hat{\mathbf{x}}_{s_j} \leftarrow I_{\xi^{\star}}\big(s_j; E_{\eta^{\star}}(\mathbf{X})\big),
    \;\;
    s_j = \tau - \sum_{i=j}^{h} \delta_i,
    \;\;
    \delta_i \sim \logU(\delta_{\min}, \delta_{\max}),
\end{equation}
so that $s_1 < \dots < s_h < \tau$ and the buffer is ordered from the oldest to the most recent point. The gaps $\delta_i$ are drawn independently, and the two bounds are set from the inter-observation gaps of the training set: $\delta_{\min}$ is the smallest gap and $\delta_{\max}$ three times the median gap. We use the median rather than the largest gap to avoid extreme outliers, which would place buffer points outside the sampling statistics of the data. The log-Uniform between these two bounds spans several orders of magnitude, the model sees both tightly and widely spaced histories, whereas on-grid histories only contain the gaps of the training data. This can be seen as a data augmentation along the time axis, the model sees more varieties of states for the same trajectory. This feature remains comparable to the other baselines methods as the model takes always in input sparse points with the same fixed $h$-length history. The velocity field receives exactly as many conditioning points as every method we compare against. We ablate this choice
against sampling the buffer on the observed grid in App. \ref{sec:ablation_sampler}.


An important remark should be highlighted: once training is finished, the teacher (t-INR), it is never used, inference time consists only rolling out a single velocity field.

\subsection{Variational Velocity Field} \label{sec:varvf}
Our current velocity field returns only a deterministic output, which is a poor description of a system especially when observations are sparse. Inspired by \cite{guo2025variational}, we introduce a latent variable $\mathbf{z} \in \mathbb{R}^{m}$ with $m \ll d$, inferred from the history segment using a temporal encoder, then we condition the velocity field on such latent. Such a modelling buys three buys three things at once: predictive uncertainty without committing to an SDE, re-use of the t-INR encoder, the encoder shares its backbone and
is warm-started from it, and at inference no SDE solver is involved, since sampling
$\mathbf{z}$ is what makes the forecast stochastic.

We model the distribution of the vector field variationally by introducing a variational posterior conditioned on the history segment $q_{\psi}(\mathbf{z} \mid \mathbf{X}_{t_{k-h:k}})$.
\begin{align}
    \begin{split}
\log & \; p_{\tau}^\phi(\dot{\mathbf{x}}_{\tau} \mid \mathbf{x}_{t_k}) 
    = \log \int p_{\tau}^\phi(\dot{\mathbf{x}}_{\tau},  \mathbf{z} \mid \mathbf{x}_{t_k})\,  dz \;
    = \log \int q_{\psi}(\mathbf{z} |\mathbf{X}_{t_{k-h:k}}) \frac{p_{\tau}^\phi(\dot{\mathbf{x}}_{\tau},  \mathbf{z} \mid \mathbf{x}_{t_k})}{q_{\psi}(\mathbf{z} |\mathbf{X}_{t_{k-h:k}})}\,  dz
    \end{split}
    \label{eq_marginalizing}
\end{align}
Applying Jensen's Inequality to \eqref{eq_marginalizing}, we derive a variational simulation-free matching objective similarly to \cite{guo2025variational}:
\begin{align}
\label{eq:ELBO}
    \log p_{\tau}^\phi(\dot{\mathbf{x}}_{\tau} \mid \mathbf{x}_{t_k}) & \ge \mathbb{E}_{q_{\psi}(\mathbf{z} |\mathbf{X}_{t_{k-h:k}})}[\log p_{\tau}^\phi(\dot{\mathbf{x}}_{\tau} \mid  \mathbf{z}, \mathbf{x}_{t_k})] -\text{KL}\left[q_{\psi}(\mathbf{z} \mid \mathbf{X}_{t_{k-h:k}}) \Vert p(\mathbf{z})\right],
\end{align}
where $p(\mathbf{z})=\mathcal{N}(\vect 0, \mathbf{I})$, we report the full derivation in App. \ref{app:elbo-derivation}.

Taking a Gaussian likelihood for the velocity match, and substituting the teacher's target \eqref{eq:inr-target}, the
bound in \eqref{eq:ELBO} gives the following training objective:
\begin{equation}
\resizebox{0.9\textwidth}{!}{%
$\begin{aligned}
\mathcal{L}_{\text{VF-INR}}(\phi,\psi) &= 
\mathbb{E}_{\substack{\pi(\mathbf{X}) \\ \tau \sim \mathcal{U}(t_0, t_T) \\ \hat{\mathbf{H}}_{\tau} \sim \pi_{\mathrm{buf}}(\cdot \mid \tau)}} \left[
\mathbb{E}_{q_{\psi}(\mathbf{z} \mid \hat{\mathbf{H}}_{\tau})}
\left[
\| v_{\phi}(\hat{\mathbf{x}}_\tau, \mathbf{z}) - \hat{\mathbf{u}}_\tau \|^2
\right]
+ \beta \mathrm{KL}\big(q_{\psi}(\mathbf{z} \mid \hat{\mathbf{H}}_{\tau}) \,\|\, p(\mathbf{z})\big)
\right],
\end{aligned}$
}
\end{equation}
where $\pi_{\mathrm{buf}}$ is the buffer distribution of \eqref{eq:history_buffer}, and the query state $\hat{\mathbf{x}}_\tau$, the target velocity $\hat{\mathbf{u}}_\tau$ and the sparse history buffer $\hat{\mathbf{H}}_{\tau}$ are given from the frozen t-INR
as described in
\S\ref{sec:teacher}. The KL term is annealed over training.

\paragraph{Inference.} The INR teacher is never used (discarded) and the forecast is an ODE rollout of
$v_{\phi}$ from the test observed $h$ history states, where repeated draws of $\mathbf{z} \sim q_{\psi}$ give the predictive ensemble, so no SDE solver is involved.
See App.~\ref{app:algorithms} for inference details.

\section{Experiments} \label{sect:exps}
\label{sec:ode_datasets}
We evaluate five systems and a real experiment on weather forecasting. Two systems are low-dimensional and non-linear: the Lorenz attractor, chaotic with strong curvature, and the RLC circuit, a 2-D oscillatory dynamic used in biology \citep{hodgkin1952} and in electrical engineering.
Three systems are spatio-temporal PDEs on a $128\times128$ grid: Reaction-Diffusion and Shallow Water from PDEBench \citep{takamoto2023pdebench}, and Gray-Scott, a chaotic and non-linear system on $64\times64$ grid, which we additionally observe on an irregular spatial grid.
We subsample observations at irregular intervals with different density ratio, $12.5\%$ (very sparse), $25\%$ (sparse), $50\%$ (dense) and $100\%$ (very dense). At $50\%$ observation density, the sampling frequency is already $4$--$13\times$ the Nyquist rate across systems (see App.~\ref{app:nyquist}), corresponding to a dense-observation regime. Full details about each experiment, dataset, models architecture and hyperparameters are in App. \ref{app:exp}-\ref{app:models}.

\textbf{Baselines.} All baselines share the formulation of \eqref{eq:prob-path} and differ in the interpolant: \textbf{TFM} \citep{zhang2024tjfm} (linear), \textbf{LFM}
\citep{islam2026longitudinal} (quadratic), \textbf{MMFM} \citep{rohbeck2025,
lee2025multimarginal} (cubic spline) and \textbf{CFO} \citep{hou2025cfo} (quintic spline). Crucially, every method uses the same velocity-field network with the same width, depth and activation, so differences are attributable to the target rather than to capacity; our method carries the additional encoder, which is warm-started from the t-INR. 

\textbf{Evaluation}. During test time, the initial condition is set after $h$ irregularly timed historical observations. The forecast is evaluated at 100\% density ratio, over a long rollout horizon to assess the error accumulation. We report the relative $L_2$ error of the rollout against the dense ground truth, cumulated from the start of the forecast over the first third (Early), two thirds (Mid) and the whole horizon (Full). We also report the Continuous Ranked Probability Score (CRPS) of the predictive ensemble, except for CFO which is a deterministic model. More details are reported in App. \ref{app:eval}

\paragraph{Sparse observation forecasting}
\label{sec:main_results}
Table~\ref{tab:main} reports the main comparison in the sparse regime, at $25\%$ density. In this setting, our approach is best performing across the systems, and the results gap increases with the curvature of the dynamics relative to the sampling. The advantage is greatest when curvature effects are strongest, for example on Lorenz ($0.25$ vs $0.89$), SWE ($0.02$ vs $0.06$) and RD ($0.08$ vs $0.16$). It is smallest for RLC ($0.38$, similar to CFO), a damped, driven oscillator whose path between two observations is well-approximated by a low-order polynomial. Same ranking is observed on CRPS. We also note that the baseline methods change their relative ranking depending on the system.

\begin{table*}[t]
\centering
\renewcommand{\arraystretch}{0.7}
\small
\caption{Forecasting performance at \textbf{sparse} observation density ($25\%$). We report full horizon CRPS and relative $L_2$ error over the full, middle, and early rollout horizons. Relative $L_2$ is normalized per-window; values are therefore not directly comparable across horizons.
}
\label{tab:main}
\resizebox{0.95\textwidth}{!}{%
\begin{tabular}{lcccc}
\toprule
Method & CRPS (Full) $\downarrow$ & Rel.\ $L_2$ (Full) $\downarrow$ & Rel.\ $L_2$ (Mid) $\downarrow$ & Rel.\ $L_2$ (Early) $\downarrow$ \\
\midrule
\multicolumn{5}{l}{\textbf{Lorenz}} \\
\midrule
TFM (Linear) & $1.41 \pm 0.20$ & $0.89 \pm 0.04$ & $0.91 \pm 0.04$ & $0.96 \pm 0.08$ \\
LFM (Quad) & $1.61 \pm 1.07$ & $0.92 \pm 0.26$ & $1.09 \pm 0.35$ & $1.26 \pm 0.35$ \\
MMFM (Spline) & $1.27 \pm 0.28$ & $0.89 \pm 0.11$ & $0.94 \pm 0.19$ & $1.04 \pm 0.08$ \\
CFO & -- & $0.89 \pm 0.22$ & $1.02 \pm 0.28$ & $1.28 \pm 0.51$ \\
Ours  & $\mathbf{0.40} \pm 0.03$ & $\mathbf{0.25} \pm 0.09$ & $\mathbf{0.22} \pm 0.10$ & $\mathbf{0.20} \pm 0.05$ \\
\midrule
\multicolumn{5}{l}{\textbf{RLC}} \\
\midrule
TFM (Linear) & $0.41 \pm 0.02$ & $0.39 \pm 0.02$ & $0.54 \pm 0.08$ & $0.46 \pm 0.04$ \\
LFM (Quad) & $0.46 \pm 0.02$ & $0.39 \pm 0.03$ & $0.53 \pm 0.04$ & $\mathbf{0.43} \pm 0.03$ \\
MMFM (Spline) & $0.41 \pm 0.02$ & $0.42 \pm 0.02$ & $0.62 \pm 0.06$ & $0.50 \pm 0.03$ \\
CFO & -- & $\mathbf{0.38} \pm 0.01$ & $0.58 \pm 0.05$ & $0.45 \pm 0.02$ \\
Ours & $\mathbf{0.39} \pm 0.02$ & $\mathbf{0.38} \pm 0.03$ & $\mathbf{0.49} \pm 0.02$ & $\mathbf{0.43} \pm 0.01$ \\
\midrule
\multicolumn{5}{l}{\textbf{SWE}} \\
\midrule
TFM (Linear) & $0.37 \pm 0.04$ & $0.08 \pm 0.01$ & $0.06 \pm 0.01$ & $0.04 \pm 0.01$ \\
LFM (Quad) & $0.34 \pm 0.05$ & $0.08 \pm 0.01$ & $0.07 \pm 0.01$ & $0.04 \pm 0.01$ \\
MMFM (Spline) & $0.24 \pm 0.05$ & $0.06 \pm 0.01$ & $0.05 \pm 0.01$ & $0.03 \pm \text{\small $\expnumber{1}{-3}$}$ \\
CFO & -- & $0.07 \pm 0.01$ & $0.04 \pm 0.01$ & $0.02 \pm 0.01$ \\
Ours & $\mathbf{0.10} \pm 0.03$ & $\mathbf{0.02} \pm 0.01$ & $\mathbf{0.01} \pm \text{\small $\expnumber{1}{-3}$}$ & $\mathbf{0.01} \pm \text{\small $\expnumber{1}{-3}$}$ \\
\midrule
\multicolumn{5}{l}{\textbf{RD}} \\
\midrule
TFM (Linear) & $0.17 \pm 0.03$ & $0.16 \pm 0.03$ & $0.12 \pm 0.01$ & $0.06 \pm \text{\small $\expnumber{1}{-3}$}$ \\
LFM (Quad) & $0.22 \pm 0.07$ & $0.22 \pm 0.06$ & $0.17 \pm 0.05$ & $0.08 \pm 0.03$ \\
MMFM (Spline) & $0.24 \pm 0.01$ & $0.24 \pm 0.01$ & $0.20 \pm 0.02$ & $0.10 \pm 0.03$ \\
CFO & -- & $0.18 \pm 0.05$ & $0.13 \pm 0.04$ & $0.06 \pm 0.02$ \\
Ours & $\mathbf{0.06} \pm 0.01$ & $\mathbf{0.08} \pm \text{\small $\expnumber{1}{-3}$}$ & $\mathbf{0.07} \pm \text{\small $\expnumber{1}{-3}$}$ & $\mathbf{0.05} \pm \text{\small $\expnumber{1}{-3}$}$ \\
\bottomrule
\end{tabular}
}
\end{table*}

\begin{table*}[b]
\centering
\small
\renewcommand{\arraystretch}{0.7}
\setlength{\tabcolsep}{5pt}
\caption{Forecasting error across observation densities and OOD: we report the relative $L_2$ over the
full horizon, lower is better.}
\label{tab:density_sweep}
\resizebox{\textwidth}{!}{%
\begin{tabular}{lccccc}
\toprule
Dataset (density) & TFM (Linear) & LFM (Quad)  & MMFM (Spline)  & CFO  & Ours  \\
\midrule
\textcolor{white}{Lorenz} ($12.5\%$)   & $0.99 \pm 0.24$ & $1.00 \pm 0.35$ & $1.33 \pm 0.20$ & $1.06 \pm 0.21$ & $\mathbf{0.31} \pm 0.09$ \\
Lorenz ($50\%$)     & $0.72 \pm 0.02$ & $0.59 \pm 0.12$ & $0.59 \pm 0.01$ & $0.60 \pm 0.10$ & $\mathbf{0.29} \pm 0.14$ \\
\textcolor{white}{Lorenz} (OOD $25\%$) & $1.75 \pm 0.82$ & $0.71 \pm 0.15$ & $0.76 \pm 0.27$ & $0.77 \pm 0.17$ & $\mathbf{0.25} \pm 0.02$ \\
\midrule
\textcolor{white}{RLC} ($12.5\%$)      & $0.89 \pm 0.14$ & $0.86 \pm 0.16$ & $0.75 \pm 0.09$ & $0.80 \pm 0.07$ & $\mathbf{0.52} \pm 0.07$ \\
RLC ($50\%$)        & $0.24 \pm 0.03$ & $0.26 \pm 0.01$ & $0.24 \pm 0.01$ & $0.22 \pm 0.02$ & $\mathbf{0.14} \pm 0.01$ \\
\textcolor{white}{RLC} (OOD $25\%$)    & $0.96 \pm 0.23$ & $0.72 \pm 0.04$ & $0.73 \pm 0.09$ & $0.74 \pm 0.06$ & $\mathbf{0.55} \pm 0.05$ \\
\midrule
\textcolor{white}{SWE} ($12.5\%$)      & $0.09 \pm \text{\small $\expnumber{1}{-3}$}$ & $0.09 \pm 0.02$ & $\mathbf{0.07} \pm 0.02$ & $0.10 \pm 0.02$ & $\mathbf{0.07} \pm 0.01$ \\
SWE ($50\%$)        & $0.06 \pm 0.01$ & $0.06 \pm 0.01$ & $0.05 \pm 0.01$ & $\mathbf{0.02} \pm 0.01$ & $0.03 \pm \text{\small $\expnumber{1}{-3}$}$ \\
\textcolor{white}{SWE} (OOD $25\%$)    & $0.11 \pm 0.01$ & $0.14 \pm 0.02$ & $0.10 \pm 0.01$ & $0.10 \pm 0.04$ & $\mathbf{0.06} \pm \text{\small $\expnumber{1}{-3}$}$ \\
\midrule
\textcolor{white}{RD} ($12.5\%$)   & $0.25 \pm 0.05$ & $0.26 \pm 0.04$ & $0.26 \pm 0.05$ & $0.15 \pm 0.05$ & $\mathbf{0.13} \pm 0.03$ \\
RD ($50\%$)     & $0.20 \pm 0.05$ & $0.20 \pm 0.04$ & $0.19 \pm 0.05$ & $\mathbf{0.13} \pm 0.07$ & $\mathbf{0.13} \pm 0.03$ \\
\textcolor{white}{RD} (OOD $25\%$) & $0.10 \pm 0.02$ & $0.15 \pm 0.05$ & $0.17 \pm \text{\small $\expnumber{1}{-3}$}$ & $0.11 \pm 0.04$ & $\mathbf{0.07} \pm 0.01$ \\
\bottomrule
\end{tabular}
}
\end{table*}

\begin{figure}[t]
    \centering
    \includegraphics[width=0.9\linewidth]{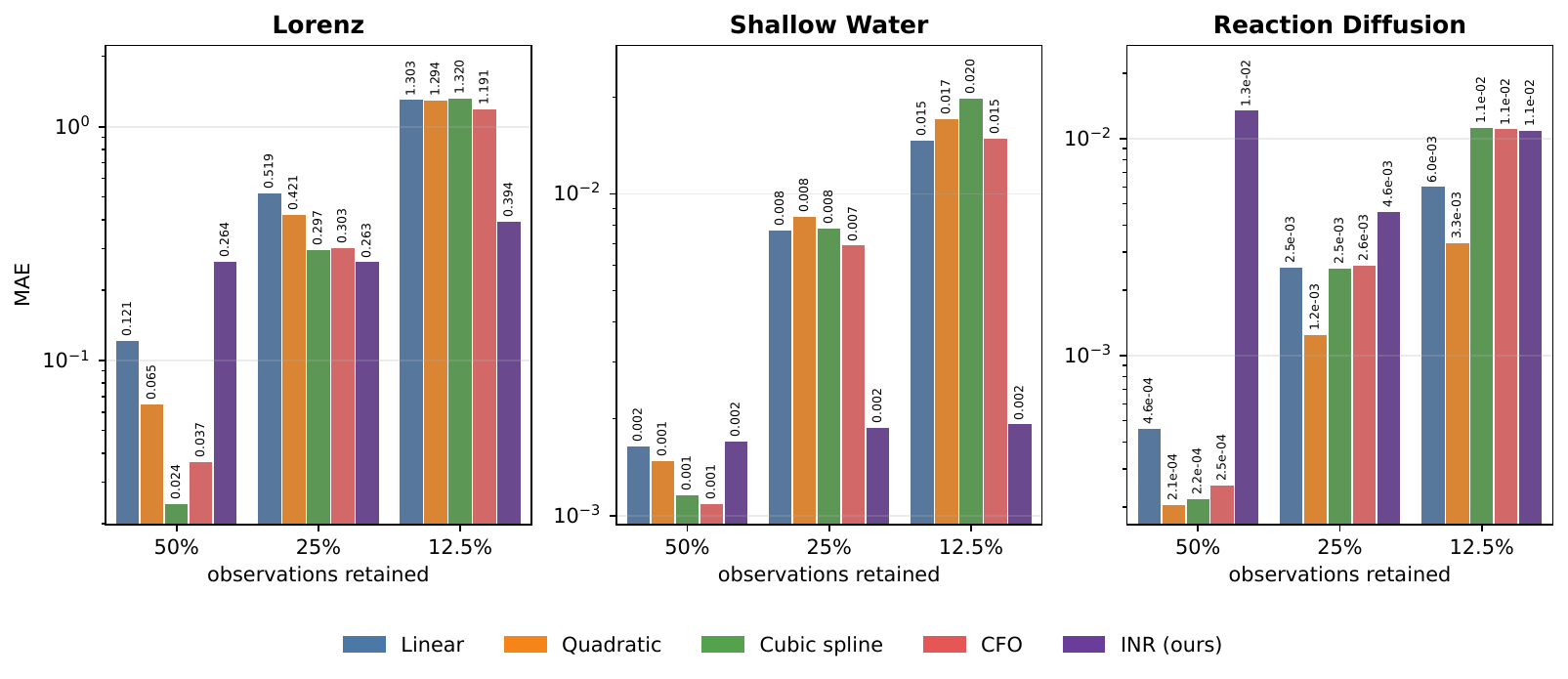}
    \caption{Interpolant reconstruction errors, measured with MAE on held-out dense frames. No fixed interpolant dominates: the ordering between linear,
    quadratic, spline and CFO changes with the system and with the retained fraction. On Lorenz 50\% and Reaction Diffusion systems the fixed interpolants
    reconstruct better than ours, yet forecast worse, see Table~\ref{tab:main}-\ref{tab:density_sweep}, what the
    velocity field regresses is the derivative of the path, not its distance to the observations.}
    \label{fig:interpolation_error}
\end{figure}

\begin{figure}[b]
  \centering
  \begin{minipage}[t]{0.32\textwidth}
    \centering
    \includegraphics[width=\textwidth]{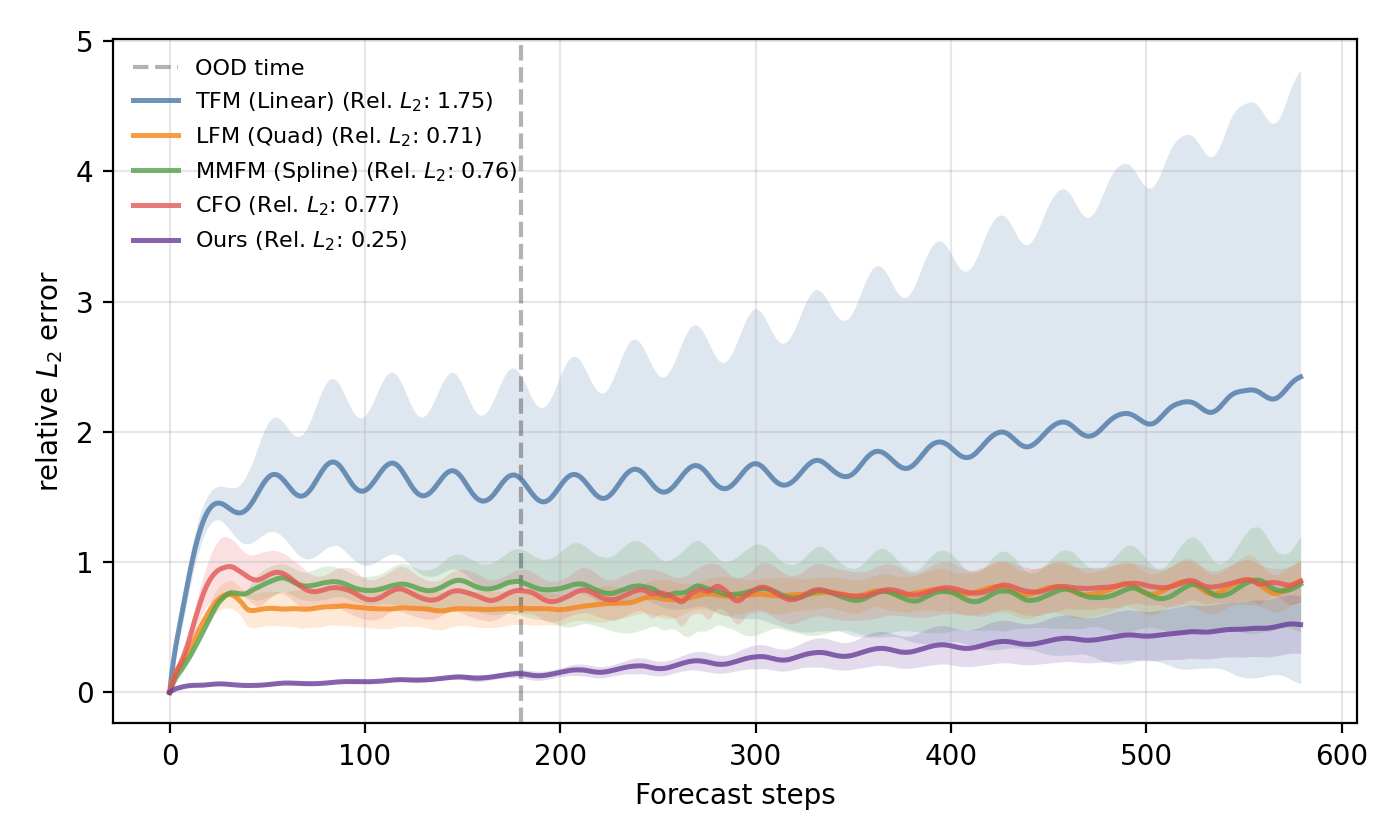}
  \end{minipage}
  \hfill 
  \begin{minipage}[t]{0.32\textwidth}
    \centering
    \includegraphics[width=\textwidth]{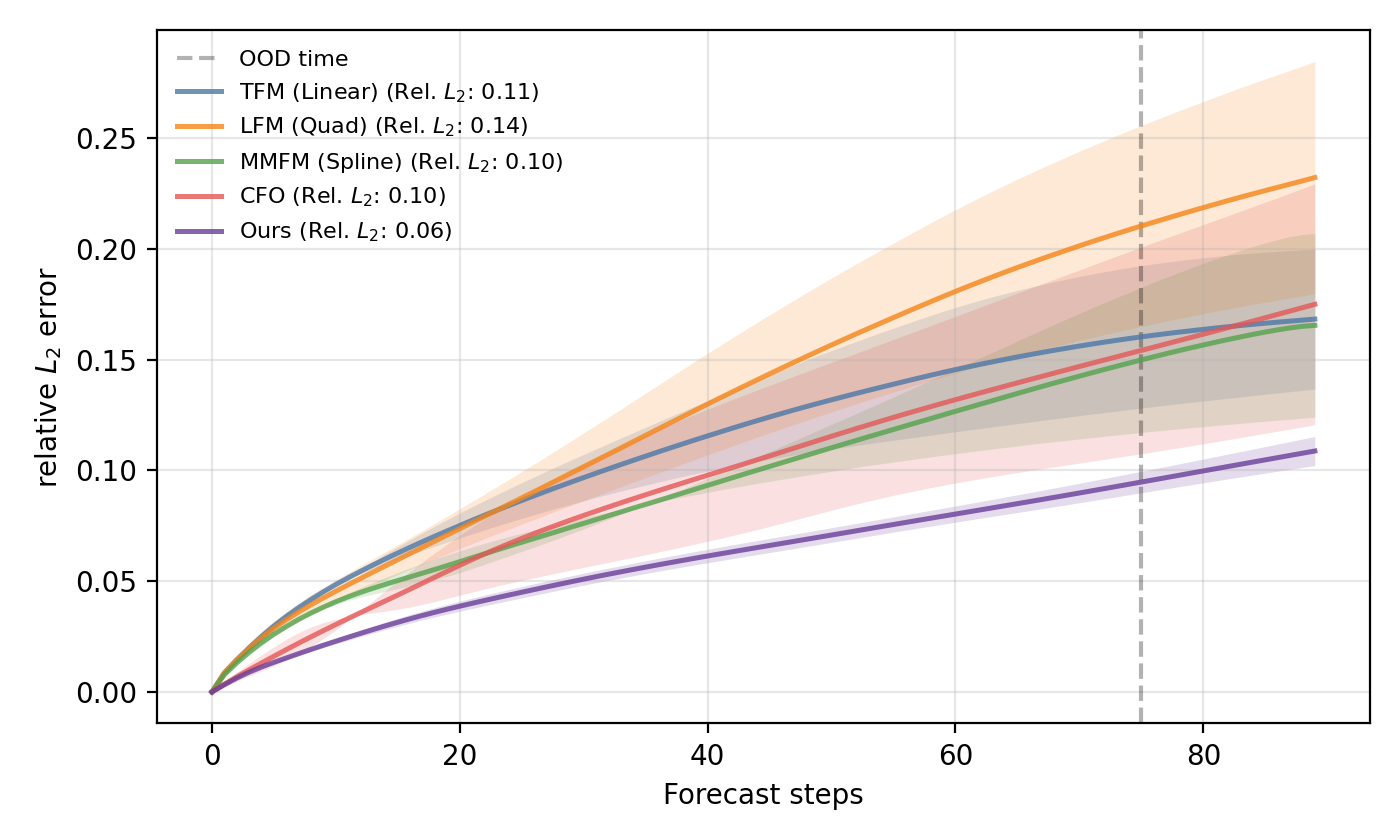}
  \end{minipage}
  \hfill 
  \begin{minipage}[t]{0.32\textwidth}
  \centering
    \includegraphics[width=\textwidth]{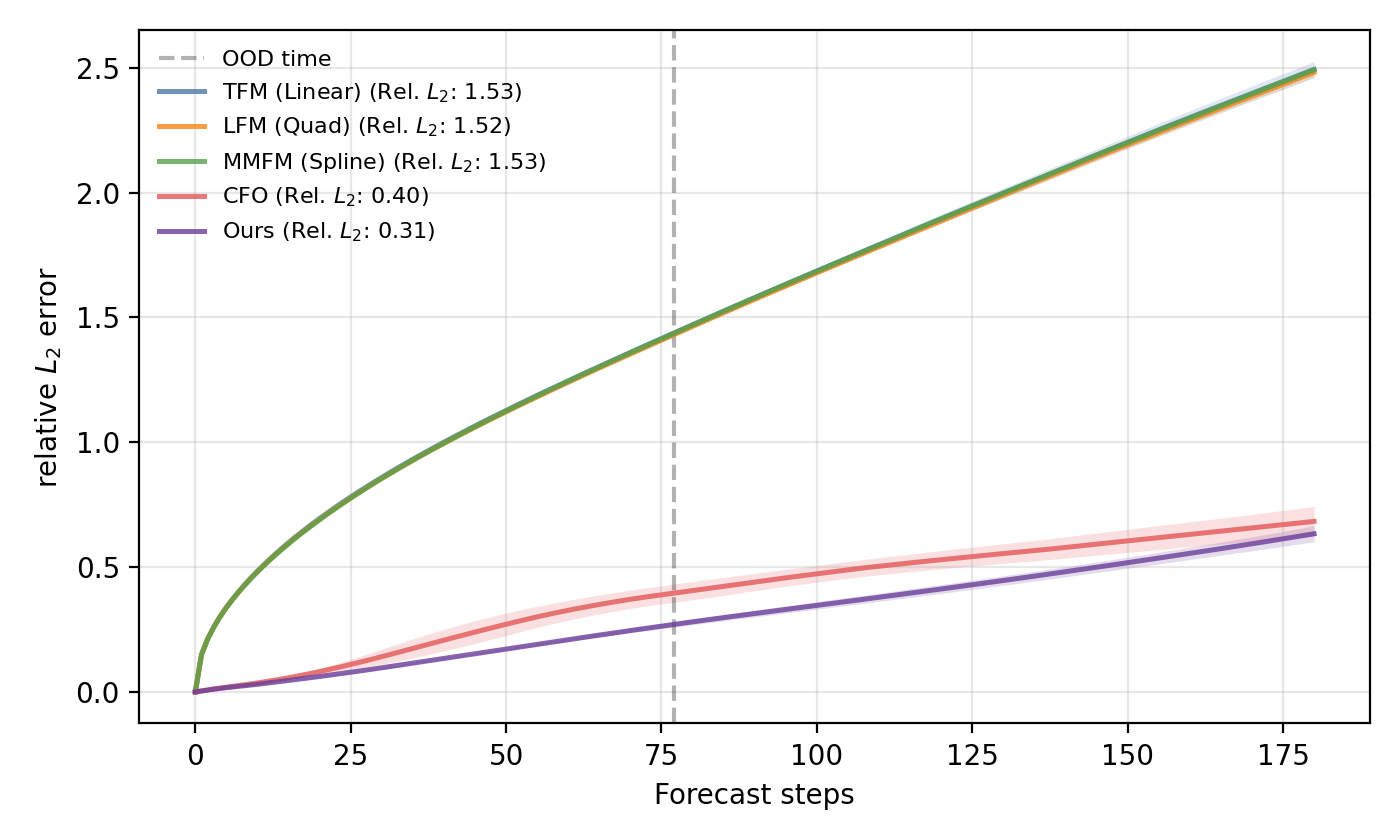}
    
  \end{minipage}
\caption{Out-of-distribution Evaluation: rollout forecast error on long-range prediction for Lorenz, Gray-Scott and SWE. The grey line marks the end of the training time horizon.}
\label{fig:ood-forecast}
\end{figure}

\paragraph{Ablating observation density, when does it help?} \label{sec:density}
Table~\ref{tab:density_sweep} varies the retained fraction of observations. At $12.5\%$ and $25\%$
we lead across the systems, particularly on Lorenz, whose dynamics exhibits the strongest curvature. At higher density ($50\%$), on SWE and RD the gap narrows, we perform on par either with MMFM or CFO. This behaviour is expected between the approaches: as intervals shrink, every reasonable interpolant converges to the same path; the regression target stops being the bottleneck. However, we notice that our method is consistent among different densities and dynamics, performing better or on par than others.

\paragraph{Interpolation vs. forecasting.} \label{sec:interp_vs_forecast}
Figure~\ref{fig:interpolation_error} measures each interpolant directly, as MAE against
held-out dense frames. No fixed interpolant dominates: the ordering among linear,
quadratic, spline and CFO changes with the system and with the retained fraction. We also highlight that although our interpolant on Lorenz at $50\%$ and Reaction-Diffusion do worse, the fixed interpolants interpolate better yet forecast
worse. The forecast models regress the time derivative of the interpolant path, not its distance to the observations (reconstruction error), and the interpolant may interpolate all observed points while exhibiting an incorrect slope between points. 

\paragraph{Out-of-distribution in time} 
We evaluate each method out of the time training range at $25\%$ density; the results are shown in Figure~\ref{fig:ood-forecast} and in the last row of each block of Table~\ref{tab:density_sweep}. We perform best on all the systems and ranking is stable out of distribution, while the baselines change order relative to one another among the experiments, highlighting that no one dominates among them. 
Figure~\ref{fig:ood-forecast} resolves the same evaluation by lead time on the non-linear systems of Lorenz, SWE and Gray-Scott:
our error grows more slowly and remains stable as the lead-time increases, whereas on Lorenz, TFM degrades sharply and drastically because in a chaotic system with strong curvature the linear interpolation is a poor choice (see visual examples in Figure \ref{fig:interp_fail_ex}). 

\paragraph{Sparsity in space: Gray-Scott} \label{sec:grayscott}
We additionally include an experiment on irregular spatial grid for the Gray-Scott dynamical system. In this experiment, we evaluate both temporal and spatial grids under the sparse setting, $25\%$ density ratio. We motivate this experiment as a robustness stress-test of the methods for evaluating their capabilities in working under adaptive resolution and grid efficiency. Although our main objective is the temporal axis (main target of the paper), our method can extend to the spatial grid by including the $\zeta \in \Omega \subset \mathbb{R}^{s}$  spatial coordinate in the t-INR, i.e $I_{\xi}\big(\tau, \zeta; E_{\eta}(\mathbf{X})\big)$. Table \ref{tab:gray-scott} reports similar trends of Table \ref{tab:main}, our method surpass all of them, while CFO is best among the baselines. The gap is largest on CRPS, where we improve by more than a factor of five with SDE baselines.

\begin{table}[t]
\centering
\renewcommand{\arraystretch}{0.7}
\small
\caption{Forecast performance Gray-Scott at $25\%$ density ratio (temporal and spatial). }
\label{tab:gray-scott}
\resizebox{\textwidth}{!}{%
\begin{tabular}{lccccc}
\toprule
Metric $\downarrow$ & TFM (Linear) & LFM (Quad) & MMFM (Spline) & CFO & Ours  \\
\midrule
CRPS (Full) & $1.86 \pm 0.02$ & $1.85 \pm \text{\small $\expnumber{1}{-3}$}$ & $1.86 \pm \text{\small $\expnumber{1}{-3}$}$ & -- & $\mathbf{0.33} \pm 0.08$ \\
Rel.\ $L_2$ (Full) & $0.38 \pm 0.02$ & $0.36 \pm 0.01$ & $0.36 \pm \text{\small $\expnumber{1}{-3}$}$ & $0.16 \pm 0.04$ & $\mathbf{0.12} \pm 0.01$ \\
Rel.\ $L_2$ (Mid) & $0.29 \pm 0.02$ & $0.28 \pm 0.01$ & $0.28 \pm \text{\small $\expnumber{1}{-3}$}$ & $0.09 \pm 0.03$ & $\mathbf{0.07} \pm \text{\small $\expnumber{1}{-3}$}$ \\
Rel.\ $L_2$ (Early) & $0.19 \pm 0.01$ & $0.18 \pm \text{\small $\expnumber{1}{-3}$}$ & $0.18 \pm \text{\small $\expnumber{1}{-3}$}$ & $0.04 \pm 0.01$ & $\mathbf{0.03} \pm \text{\small $\expnumber{1}{-3}$}$ \\
\bottomrule
\end{tabular}
}
\end{table}

\paragraph{Real data: medium-range weather on ERA5.} \label{sec:era5} Every system was so far synthetic. We include an experiment on climate data, ERA5 \citep{weatherbench2020}. We choose this setup because ERA5 is archival rather than observational: reanalyses are dense, climate archives are often stored at different temporal resolutions for cost reasons. A common regime for earth observation and remote sensing data. Our aim is not to compete with state-of-the-art weather models, which are trained on dense
archives with far larger, domain-specific architectures; our question is narrower: with the velocity field held fixed, we test whether lifting the linear-interpolant assumption helps. We use ERA5 at $50\%$ of a $15$-day window ($30$ of $60$ frames, mean gap $12$\,h) at random irregular times, on the fixed $5.625^\circ$ grid with five variables (z500, t850, t2m, u10, v10). 

\begin{figure}[t]
    \centering
    \includegraphics[width=\linewidth]{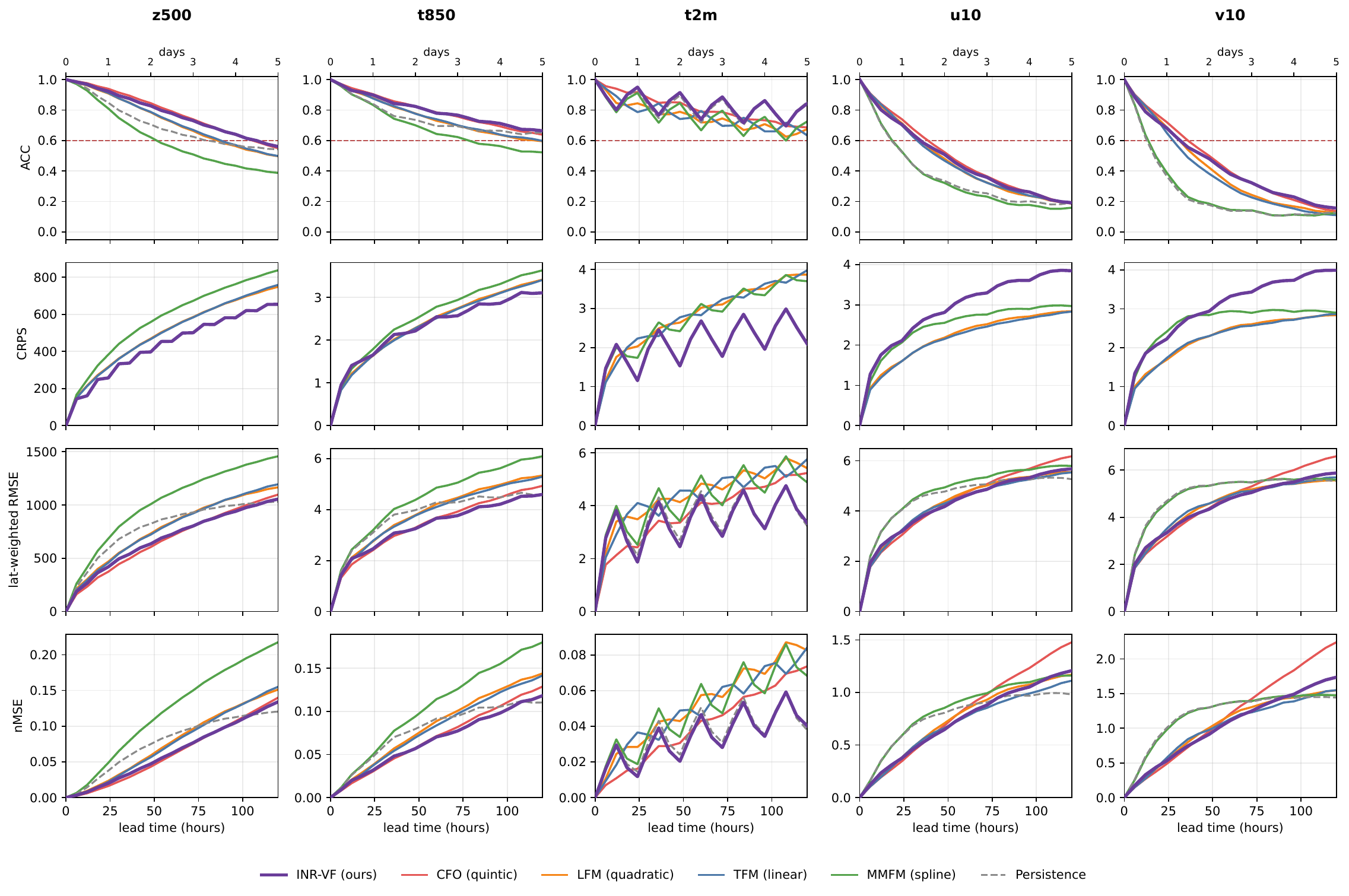}
    \caption{ERA5 skill against lead time of $120$\,h (5 days). Rows: ACC, CRPS,
    $\cos(\mathrm{lat})$-weighted RMSE, nMSE; columns: the five variables. The dashed red
    line marks $\mathrm{ACC}=0.6$, the conventional limit of a useful forecast. Persistence is the last observation carried forward.}
    \label{fig:era5_skill}
\end{figure}

From a buffer of $10$ history observations, all models generate a full rollout forecast, scored every $6$\,h with $\cos(\mathrm{lat})$-weighted RMSE, Anomaly Correlation Coefficient (ACC), CRPS,
and nMSE; Figure~\ref{fig:era5_skill} reports error against lead time up to $120$\,h
($5$-day). The linear, quadratic, and cubic interpolants lose ACC quickly when going beyond the $0.6$ threshold. This threshold is usually employed to indicate a useful forecast, while our method is considerably more robust when looking at a long time horizon. Wind components are the hardest to forecast: all methods, including ours, drop below around $36$\, h. CFO is the strongest competing baseline, matching our method closely on this metric, but ours is more consistent across all metrics, including nMSE,
where it performs better on wind and t2m. We give in App.~\ref{app:era5} the complete protocol and more results at $234$\,h and $48$\,h lead.
(Figures~\ref{fig:era5_skill_48}-~\ref{fig:era5_skill_full}).

\section{Conclusion}
Simulation-free trajectory matching is not a universal solution; its efficacy is constrained by the interpolant assumption.
In this work, we studied this limitation under sparse observations, where fixed local interpolants do not accurately represent the dynamics. Our experiments highlight that no fixed interpolant best dominates across systems or sampling densities. This indicates that such interpolants remain local approximations based on only a few observations.
We address this issue by learning a non-linear interpolant with an amortized temporal INR (t-INR), which captures both local and global structure of the observations. We incorporate the learned interpolant into a simulation-free trajectory-matching method and introduce a variational velocity field for probabilistic forecasting. In our experiments, this approach improves long-horizon rollout accuracy and stability at matched velocity-field capacity, without requiring an SDE solver. Main limitations relate to pre-train the INR, unlike fixed-interpolant baselines, and it is most beneficial for sparse observations and non-linear, curved dynamics. For dense data, linear interpolation may be adequate, at the cost of acquiring measurements at high resolution that capture velocity changes. More generally, interpolants should consider the quality of the derivatives they induce, and not only their state-reconstruction quality.

\clearpage
\subsection*{AI use statement}
In this work, we used generative AI tools to assist with implementing parts of the pipeline and baselines in the code,
also to provide feedback on and discuss the experimental design, and visualize results.
We did \textbf{not} use generative AI tools to generate synthetic datasets, develop the theoretical framework,
propose or refine hypotheses, formulate mathematical claims, contribute to any part of the derivations,
or assist in writing any proofs. Data translation, dataset cleaning and reformatting, and qualitative or
thematic data analysis were not applicable to this work. We additionally used generative AI tools to
edit and polish the manuscript for readability and conciseness, check the consistency of notation and cross-references,
write and refactor analysis and plotting code, generate figures and tables from logged experimental runs. All AI-assisted work was reviewed by the authors. Every sentence was reviewed and revised by the authors; every citation retained in the manuscript was verified
against its original source; and all AI-assisted code was verified by the authors. All reported results and
numerical values come from experiments executed by the authors using our code.
The authors take full responsibility for the final content of this work, including all text, claims, analyses,
and other artifacts produced with the assistance of generative AI.

\subsection*{Ethics statement}
Time-series tasks, including forecasting and simulating dynamics, are widely used across many scientific and engineering applications. In this work, we propose a new approach for handling observations in which measurements are sparse and irregular, reflecting more realistic real-world conditions. We evaluated the robustness of our method against several baseline approaches on a simulated dataset, following established practices. In addition, we extended our evaluation to a more practical real-world setting in weather forecasting, where temporal measurements are missing. Importantly, this approach can be applied in a broad range of domains, including weather, medicine, military applications, aerodynamics, and industry.

\subsection*{Reproducibility statement}
We have taken care to make every result in this paper reproducible. The method section and
appendix specify the architecture, the objective, pseudocode training and evaluation
steps. We report the complete set of hyper parameters in the appendix for our method and every baseline.

\clearpage
\bibliography{iclr2027_conference}
\bibliographystyle{iclr2027_conference}

\clearpage
\appendix

\section{Derivations}

\subsection{ELBO Derivation}
\label{app:elbo-derivation}
\begin{align}
\log p_{\tau}^\phi(\dot{\mathbf{x}}_{\tau} \mid \mathbf{x}_{t_k}) 
&= \log \int p_{\tau}^\phi(\dot{\mathbf{x}}_{\tau}, \mathbf{z} \mid \mathbf{x}_{t_k})\, d\mathbf{z} \\
&= \log \int q_{\psi}(\mathbf{z} \mid \mathbf{X}_{t_{k-h:k}}) \frac{p_{\tau}^\phi(\dot{\mathbf{x}}_{\tau}, \mathbf{z} \mid \mathbf{x}_{t_k})}{q_{\psi}(\mathbf{z} \mid \mathbf{X}_{t_{k-h:k}})}\, d\mathbf{z} \\
&\geq \int q_{\psi}(\mathbf{z} \mid \mathbf{X}_{t_{k-h:k}}) \log \frac{p_{\tau}^\phi(\dot{\mathbf{x}}_{\tau}, \mathbf{z} \mid \mathbf{x}_{t_k})}{q_{\psi}(\mathbf{z} \mid \mathbf{X}_{t_{k-h:k}})}\, d\mathbf{z} \quad \text{(Jensen's inequality)} \\
&= \int q_{\psi}(\mathbf{z} \mid \mathbf{X}_{t_{k-h:k}}) \log \frac{p_{\tau}^\phi(\dot{\mathbf{x}}_{\tau} \mid  \mathbf{z}, \mathbf{x}_{t_k}) p_{\tau}^\phi(\mathbf{z} \mid \mathbf{x}_{t_k})}{q_{\psi}(\mathbf{z} \mid \mathbf{X}_{t_{k-h:k}})}\, d\mathbf{z}\\
&= \mathbb{E}_{q_{\psi}(\mathbf{z} \mid \mathbf{X}_{t_{k-h:k}})}\left[\log p_{\tau}^\phi(\dot{\mathbf{x}}_{\tau}\mid  \mathbf{z}, \mathbf{x}_{t_k}) - \log \frac{q_{\psi}(\mathbf{z} \mid \mathbf{X}_{t_{k-h:k}})}{p_{\tau}^\phi(\mathbf{z} \mid \mathbf{x}_{t_k})}\right] \\
&= \mathbb{E}_{q_{\psi}(\mathbf{z} \mid \mathbf{X}_{t_{k-h:k}})}[\log p_{\tau}^\phi(\dot{\mathbf{x}}_{\tau} \mid  \mathbf{z}, \mathbf{x}_{t_k})] - \text{KL}\left[q_{\psi}(\mathbf{z} \mid \mathbf{X}_{t_{k-h:k}}) \Vert p_{\tau}^\phi( \mathbf{z} \mid \mathbf{x}_{t_k})\right]
\end{align}

Assuming independent and data and time independent priors:
\begin{equation}
   p_{\tau}^\phi( \mathbf{z} \mid \mathbf{x}_{t_k}) = p(\mathbf{z}) 
\end{equation}
The objective becomes:
    \begin{align}
        \log p_{\tau}^\phi(\dot{\mathbf{x}}_{\tau} \mid \mathbf{x}_{t_k}) &\geq \mathbb{E}_{q_{\psi}(\mathbf{z} |\mathbf{X}_{t_{k-h:k}})}[\log p_{\tau}^\phi(\dot{\mathbf{x}}_{\tau} \mid  \mathbf{z}, \mathbf{x}_{t_k})]  -\text{KL}\left[q_{\psi}(\mathbf{z} \mid \vect{X}_{k-h:k}) \Vert p(z)\right],
    \end{align}

\section{Method Details} \label{app:algorithms}
This section summarises the training and the inference procedures of our method. Algorithm~\ref{alg:train} describes the training of the variational velocity field, and assumes that the t-INR of \S\ref{sec:tinr} is already trained and frozen, since the pre-training of the t-INR only involves a simple reconstruction loss of \S\ref{sec:tinr}. Algorithm~\ref{alg:infer} describes the forecast at inference time, where no t-INR is involved: the rollout only uses the encoder and the velocity field.

\begin{algorithm}[H]
\DontPrintSemicolon
\caption{Training the variational velocity field from the frozen t-INR teacher}
\label{alg:train}
\KwIn{$\mathcal{D}$; frozen t-INR $(\xi^{\star},\eta^{\star})$; $h$; $\delta_{\min},\delta_{\max}$; $\beta$}
\KwOut{$v_{\phi}$, $q_{\psi}$}
warm-start $\psi$ from $E_{\eta^{\star}}$\;
\Repeat{converged}{
  $\mathbf{X}\sim\mathcal{D}$ ,\; $\mathbf{Z}\leftarrow E_{\eta^{\star}}(\mathbf{X})$ ,\; $\tau\sim\mathcal{U}(t_0,t_T)$\;
  $\hat{\mathbf{x}}_{\tau},\hat{\mathbf{u}}_{\tau} \leftarrow
     I_{\xi^{\star}}(\tau;\mathbf{Z}),\, \partial_{\tau}I_{\xi^{\star}}(\tau;\mathbf{Z})$
     \tcp*{\eqref{eq:inr-target}}
  $\delta_{1:h}\sim\logU(\delta_{\min},\delta_{\max})$ ,\;
  $s_j\leftarrow\tau-\textstyle\sum_{i=j}^{h}\delta_i$\;
  $\hat{\mathbf{H}}_{\tau}\leftarrow I_{\xi^{\star}}(s_{1:h};\mathbf{Z})$
     \tcp*{\eqref{eq:history_buffer}}
  $\mathbf{z}\sim q_{\psi}(\cdot\mid\hat{\mathbf{H}}_{\tau})$ \tcp*{reparametrised}
  $\mathcal{L}\leftarrow
     \lVert v_{\phi}(\hat{\mathbf{x}}_{\tau}\mid\mathbf{z})-\hat{\mathbf{u}}_{\tau}\rVert^{2}
     +\beta\,\mathrm{KL}(q_{\psi}\Vert p)$\;
  $(\phi,\psi)\leftarrow$ gradient step on $\mathcal{L}$\;
}
\Return $v_{\phi},q_{\psi}$ \tcp*{discard the t-INR}
\end{algorithm}

\begin{algorithm}[H]
\DontPrintSemicolon
\caption{Forecasting at inference time: rolling encoder, no t-INR}
\label{alg:infer}
\KwIn{observed buffer $\mathbf{H}=(\mathbf{x}_{t_{k-h}},\dots,\mathbf{x}_{t_{k-1}})$ ,\;
      $\mathbf{x}_{r_0}=\mathbf{x}_{t_k}$ ,\; forecast times $r_0<r_1<\dots<r_M$ ,\;
      trained $v_{\phi},q_{\psi}$}
\KwOut{forecast $\mathbf{x}_{r_1:r_M}$}
\For{$i=1$ \KwTo $M$}{
  $\mathbf{z}\sim q_{\psi}(\cdot\mid\mathbf{H})$ \tcp*{re-encode the current buffer}
  $\mathbf{x}_{r_i}\leftarrow \mathbf{x}_{r_{i-1}}
     +\int_{r_{i-1}}^{r_i} v_{\phi}(\cdot\mid\mathbf{z})\,dt$ \tcp*{ODE step, no SDE solver}
  $\mathbf{H}\leftarrow(\mathbf{H}_{2:h},\mathbf{x}_{r_{i-1}})$ \tcp*{FIFO: drop the oldest, append}
}
\Return $\mathbf{x}_{r_1:r_M}$ \tcp*{fresh $\mathbf{z}$ per ensemble member}
\end{algorithm}

\paragraph{Buffer update at inference.} For readability, Algorithm~\ref{alg:infer} updates the history buffer after every integration step. In practice, the buffer is updated on a \emph{macro step} $\Delta_{\text{macro}}$, set to the median inter-observation gap of the training set, and the encoder is re-evaluated after each update. The forecast is evaluated on the dense grid, and refreshing the buffer at every step of the ODE solver would present the velocity field with histories far denser than the ones seen during training. The macro step is used for our method and for its on-grid ablation (App. ~\ref{sec:ablation_sampler}). In the ablations, all the baselines are also rolled out with the same macro-step update, so that every method receives its history buffer at the same cadence. The macro step is most relevant in the sparse and very sparse settings, where the gap between the training observations and the dense forecast grid is the largest.

\paragraph{Computing the target velocity.} \label{app:autodiff} The target velocity $\hat{\mathbf{u}}_\tau = \partial_\tau I_{\xi^{\star}}(\tau; E_{\eta^{\star}}(\mathbf{X}))$ of \eqref{eq:inr-target} is the derivative of the decoder output with respect to a single scalar input, the query time $\tau$. The encoder output $E_{\eta^{\star}}(\mathbf{X})$ does not depend on $\tau$, so it is computed once per trajectory and reused for all the query times, and only the decoder is differentiated. Since the t-INR is frozen, no computational graph is kept: the target is evaluated without tracking gradients with respect to the parameters, and enters the loss as a constant. We use two implementations, depending on the dimension of the state.
For the low-dimensional systems, we use reverse-mode automatic differentiation: for each state dimension, we backpropagate the sum of that dimension over all the query times to $\tau$. Since each query only depends on its own time, this gives the exact derivative of every query in one backward pass per state dimension, i.e., two or three passes.
For the spatio-temporal systems and ERA5, the state has $C \times H \times W$ dimensions, and reverse mode would require one backward pass per dimension. We instead use forward-mode automatic differentiation, i.e., a Jacobian-vector product with a unit tangent on $\tau$, which returns the state and its exact time derivative on the whole field in a single forward evaluation, at a cost of about two decoder passes.
In both cases, the t-INR operates on min-max normalised time and states, and the target is expressed in physical units by the chain rule through the two normalisations, $\hat{\mathbf{u}}_\tau = \frac{x_{\max}-x_{\min}}{t_{\max}-t_{\min}}\, \partial_{\tilde{\tau}}\, \tilde{I}_{\xi^{\star}}(\tilde{\tau})$, where the tilde denotes normalised quantities. In reverse mode, this factor is included automatically by differentiating through the normalisations; in forward mode, it is applied explicitly to the output of the Jacobian-vector product.

\section{Experimental Setup}
\label{app:exp}
We evaluate five synthetic systems and a real-world weather forecasting task, which is described separately in App. ~\ref{app:era5}. Two systems are low-dimensional and non-linear: the Lorenz attractor, chaotic with strong curvature, and the RLC circuit, a 2-D oscillatory system. Three systems are spatio-temporal PDEs: Reaction-Diffusion and Shallow Water, following the work of \citet{hou2025cfo} and downloading them from PDEBench \citep{takamoto2023pdebench} on a $128\times128$ grid, and Gray-Scott on a $64\times64$ grid, which we additionally observe on an irregular spatial grid.

\subsection{Datasets} \label{app:datasets}
For every system, the trajectories are first simulated on a dense and regular time grid, which defines the $100\%$ density ratio. The training observations are then obtained by retaining $12.5\%$, $25\%$ or $50\%$ of the dense grid at irregular times, drawn independently for each trajectory. The validation and test trajectories uses the full resolution, and are subsampled only at evaluation time (App. ~\ref{app:eval}).

\paragraph{RLC.} The RLC series circuit, consisting of a resistor (R), capacitor (C), and inductor (L), models a broad class of transfer functions. The system state at time $t$ is defined as $\mathbf{x}_t = [I_t, U_t]^\top$, where $U_t$ denotes the capacitor voltage and $I_t$ the circuit current. The state evolution follows:
\begin{align}
\frac{\mathrm{d}}{\mathrm{d}t} \begin{bmatrix} I_t \\ U_t \end{bmatrix} = \begin{bmatrix} \frac{1}{L}(V_t - U_t) \\ \frac{I_t}{C} \end{bmatrix} + \begin{bmatrix} -\frac{R}{L}I_t \\ 0 \end{bmatrix}.
\end{align}
The dataset is generated by sampling $L$ and $C$ uniformly from $[1, 3] \times [0.5, 1.5]$, and $R \sim \mathcal{U}(1, 3)$. The input voltage $V_t$ is a fixed, known signal. Initial conditions are given by $U_0 \sim \mathcal{N}(0, 1)$ and $I_0 = 0$. Each trajectory is simulated from $t_0 = 0$ to $t_{\text{max}} = 10$ seconds with a time resolution of $\Delta t = 0.05$\,s, yielding $200$ time points at $100\%$ density ratio per trajectory. The training, validation, and test sets contain respectively $500$, $125$ and $125$ trajectories.

\paragraph{Lorenz.} We analyse the Lorenz system for varying initial conditions and parameters. The system dynamics are governed by:
\begin{align}
\frac{\mathrm{d}u}{\mathrm{d}t} &= \sigma (v - u), \nonumber \\
\frac{\mathrm{d}v}{\mathrm{d}t} &= u (\rho - w) - v, \nonumber \\
\frac{\mathrm{d}w}{\mathrm{d}t} &= u v - \beta w .
\end{align}
We generate the trajectories with $u(0), v(0), w(0)$ sampled from a Gaussian distribution, and the parameters are also randomly sampled with $\sigma \sim \mathcal{U}(9.5, 10.5)$, $\rho \sim \mathcal{U}(27.0, 29.0)$ and $\beta \sim \mathcal{U}(2.6, 2.8)$. Each trajectory spans the time interval $t_0=0$ to $t_{\text{max}}=4$ with a time resolution of $\Delta t=0.02$, which gives $200$ states per trajectory at $100\%$ density ratio. The training set is made of $500$ trajectories, while the validation and test sets consist of $100$ trajectories each.

\paragraph{Shallow Water (PDEBench).} We use the 2-D radial dam-break problem of PDEBench \citep{takamoto2023pdebench}, whose state is the water height on a $128\times128$ grid. Each trajectory is stored with $100$ frames, of which we keep the first $90$ as the dense grid at $100\%$ density ratio. The $1000$ trajectories are shuffled and split into $800$, $100$ and $100$ trajectories for training, validation and test.

\paragraph{Reaction-Diffusion (PDEBench).} We use the 2-D diffusion-reaction problem of PDEBench \citep{takamoto2023pdebench}, which couples two non-linear fields, an activator and an inhibitor, on a $128\times128$ grid. As for Shallow Water, we keep the first $90$ of the $100$ stored frames, and split the $1000$ trajectories into $800$, $100$ and $100$ trajectories.

\paragraph{Gray-Scott.} The Gray-Scott model describes two species $u$ and $v$ that react and diffuse,
\begin{align}
\partial_t u = \nu_u \Delta u - u v^2 + F(1-u), \qquad
\partial_t v = \nu_v \Delta v + u v^2 - (F+k)\, v .
\end{align}
We simulate it with a spectral solver on a $64\times64$ periodic grid, with $\nu_u = 2\cdot10^{-5}$ and $\nu_v = 10^{-5}$, and initial conditions given by random Gaussian blobs. The feed and kill rates are sampled for each trajectory as $F \sim \mathcal{U}(0.030, 0.045)$ and $k \sim \mathcal{U}(0.060, 0.064)$. Each trajectory has $101$ frames at $\Delta t = 1$. The training, validation and test sets contain $500$, $125$ and $125$ trajectories. In addition to the temporal subsampling, each observed frame retains $25\%$ of its pixels, at random positions that change from frame to frame.

\paragraph{Out-of-distribution test sets.} To evaluate the forecast beyond the time range seen during training, we build for each system an additional test set whose trajectories extend past the training horizon, while the physical parameters and the initial conditions follow the same distributions as in training.
For Lorenz, we regenerate the $100$ test trajectories with the same random seed, and integrate them up to $t_{\text{max}} = 12$, three times the training span ($600$ dense points at $\Delta t = 0.02$), retaining $150$ of them ($25\%$). The in-distribution test trajectories are thus the first third of the out-of-distribution ones.
For RLC, the $125$ test circuits are simulated in the same way up to $t_{\text{max}} = 30$\,s, three times the training span ($598$ dense points), retaining $150$ of them ($25\%$).
For Shallow Water and Reaction-Diffusion, the PDEBench trajectories contain $100$ frames, of which only the first $90$ are used for training. The out-of-distribution test set uses the same $100$ test trajectories over all their $100$ frames, so the final frames lie beyond the training time range, and $22$ frames are retained at irregular times.
For Gray-Scott, the $125$ test trajectories are simulated with the same seed for twice as long ($201$ frames over $[0, 3000]$), retaining $25$ observations.
In all cases, the history and the initial condition are taken as in the in-distribution evaluation (Appendix~\ref{app:eval}), and the forecast is rolled out until the end of the extended trajectory, so that its later part lies outside the time range of the training data.

\subsection{Evaluation protocol} \label{app:eval}
All methods are evaluated on the same held-out test trajectories, which are available at the full ($100\%$) temporal resolution. For each trajectory, the history is formed by $h$ \textbf{sparse} observations taken from a stride of the dense grid, chosen to match the sparsity of the training data. We use $h = 10$ in the sparse and dense settings and $h = 5$ in the very sparse setting, identically for every method, including the spatio-temporal experiments. The initial condition is the observation that immediately follows the history, and the forecast is then evaluated on every remaining point of the dense grid.

We report the relative $L_2$ error of the rollout against the dense ground truth, accumulated from the start of the forecast over the first third (Early), the first two thirds (Mid) and the whole horizon (Full) of the scored window. The probabilistic methods are additionally evaluated with the Continuous Ranked Probability Score (CRPS), computed from an ensemble of $20$ forecasts: the SDE baselines draw fresh diffusion noise for each member, while our method draws a new latent $\mathbf{z} \sim q_{\psi}$. CFO is a deterministic model, and its CRPS is therefore not reported. For the out-of-distribution evaluation, the forecast starts from initial times that lie outside the time range seen during training. All results are reported as mean and standard deviation over four training seeds for the small dimension problems (RLC, and Lorenz) and three training seeds for the high-dimensional problems.

\subsection{Temporal bandwidth of the synthetic systems -- Nyquist rate}
\label{app:nyquist}
The density labels used in the experiments ($100\%$, $50\%$, etc.) are fractions of the
simulation's time grid. To make these labels meaningful, we measure the temporal
bandwidth of each system and re-express the densities in the unit that the
Nyquist--Shannon theorem uses: samples per cycle of the fastest-varying component of the
trajectory.

\textbf{Evaluation.} For each system, we take the dense trajectories and, for every state
variable, look at how its value fluctuates over time (subtracting its average so we isolate
the fluctuations). We compute the frequency content of these fluctuations and average the
result across trajectories, and for the PDE systems, across $256$ randomly chosen spatial
locations, since each point in space evolves as its own time series. From this, we extract
$f_q$: the frequency below which $99\%$ of the signal's variation is contained, i.e., the
fastest frequency that meaningfully matters. We convert $f_q$ into samples per cycle,
\begin{equation}
    \text{samples per cycle}(d) \;=\; \frac{d}{f_q \, \Delta t_{\text{dense}}},
\end{equation}
where $d$ is the retained fraction of the grid. By the Nyquist-Shannon theorem, at least
$2$ samples per cycle are needed to resolve a signal; below that, the corresponding dynamics
can no longer be recovered from the samples.

\begin{table}[h]
\centering
\caption{Samples per cycle of the fastest relevant frequency ($f_q$), for each retained
fraction of the time grid. At $50\%$, every
system is still oversampled $4$--$13\times$; at $100\%$, $8$--$25\times$ -- which is why we
call these two budgets dense and very dense.}
\label{tab:nyquist}
\begin{tabular}{lcccc}
\toprule
System & $100\%$ & $50\%$ & $25\%$ & $12.5\%$ \\
\midrule
Lorenz & 15.4 & 7.7 & 3.9 & 1.9 \\
RLC & 50.5 & 25.2 & 12.6 & 6.3 \\
SWE & 20.6 & 10.3 & 5.2 & 2.6 \\
Reaction-Diffusion & 47.6 & 23.8 & 11.9 & 6.0 \\
Gray-Scott & 21.7 & 10.8 & 5.4 & 2.7 \\
\bottomrule
\end{tabular}
\end{table}

\textbf{Results.} Table~\ref{tab:nyquist} makes one point clearly: our whole sweep is
already in the dense regime. At $50\%$, every system keeps $4$--$13\times$ more samples
than needed to resolve its dynamics; at $100\%$, $8$--$25\times$ more. So going from $50\%$
to $100\%$ adds no new information about the dynamics, only redundant samples, which in
practice are exactly the samples that are expensive or hard to obtain.

Second, the same percentage does not mean the same thing in every system. At $25\%$ Lorenz
keeps $3.9$ samples per cycle while RLC keeps $12.6$, so the two systems are in rather
different regimes despite carrying the same label. It is useful to note that this ordering
follows the ordering of our margins in Table~\ref{tab:main}: we gain the most where the
sampling is closest to the floor, and the least where the grid is still comfortable.

Finally, only Lorenz at $12.5\%$ actually drops below the floor, at $1.9$ samples per cycle.
We flag this as a possible undersampled case, but only as a warning sign, not as proof of
aliasing. Our actual point is the opposite: fixed interpolants already lose accuracy at
$50\%$, where by this measure sampling is more than sufficient.

\section{Models and Hyperparameters} \label{app:models}

\subsection{t-INR} \label{app:tinr}
The t-INR is pre-trained once per system, on the training observations only, with the reconstruction loss of \S\ref{sec:tinr}. We use AdamW with a learning rate of $10^{-4}$, and select the checkpoint with the lowest reconstruction error on the validation set. We use two architectures, depending on the dimension of the state.

\paragraph{Low-dimensional systems (Lorenz, RLC).} The encoder $E_{\eta}$ is a lightweight Transformer ($3$ layers, $8$ heads, width $256$) over the observed pairs $(t_k, \mathbf{x}_{t_k})$, which returns a latent sequence of dimension $64$. The decoder $I_{\xi}$ embeds the query time with a SIREN ($4$ layers, width $128$, $\omega_0 = 10$) and attends to the latent sequence with cross-attention to predict the state.

\paragraph{Spatio-temporal systems (Shallow Water, Reaction-Diffusion, Gray-Scott).} The encoder combines factorised (2+1)-D convolutional blocks, which downsample each frame, with a Transformer ($4$ layers, $8$ heads, width $256$) along time. The decoder queries this sequence with the same SIREN cross-attention ($\omega_0 = 10$), across all the coordinates, and maps the resulting context vector to the full field with a convolutional decoder that mirrors the encoder, with circular padding. Reconstruction loss is computed on the observed pixels.

\paragraph{Segments.} A spatio-temporal trajectory can be too large to be encoded at once within the memory of a single GPU. As is common practice, the t-INR is therefore trained on segments of consecutive observations: the encoder takes one segment as input, and the decoder interpolates within the time span of that segment. We use segments of $22$ observations for Shallow Water and Reaction-Diffusion, $20$ for Gray-Scott and $30$ for ERA5; a trajectory with more observations is split into consecutive segments. The history buffer of the velocity field ($h = 10$ observations) remains shorter than a segment.

\subsection{Velocity field} \label{app:vf}
\paragraph{Low-dimensional systems.} The velocity field $v_{\phi}$ is an MLP with four hidden layers of width $128$ and SELU activations, and the query time is embedded with $10$ Fourier frequencies.

\paragraph{Spatio-temporal systems.} The velocity field is a residual convolutional network ($32$ base channels, $2$ residual blocks per stage) that maps the state field to the velocity field. The time embedding and the conditioning vector are combined by an MLP of width $256$, and injected at every stage by feature-wise linear modulation (FiLM).

For all systems, the posterior $q_{\psi}$ shares the backbone of the t-INR encoder, from which it is warm-started, and outputs the mean and the log-variance of $\mathbf{z}$.

\subsection{Baselines} \label{app:baselines}
All baselines share the formulation of \eqref{eq:prob-path} and differ only in the interpolant: \textbf{TFM} \citep{zhang2024tjfm} (linear), \textbf{LFM} \citep{islam2026longitudinal} (quadratic), \textbf{MMFM} \citep{rohbeck2025, lee2025multimarginal} (cubic spline) and \textbf{CFO} \citep{hou2025cfo} (quintic spline). They use the same velocity-field network, the same history length $h$ and the same optimisation settings as our method (Table~\ref{tab:hparams}), so that differences are attributable to the regression target rather than to capacity. TFM, LFM and MMFM additionally learn the diffusion term $g_{\phi}$ of \eqref{eq:TFM} with a small separate network: an MLP with two hidden layers of width $64$ for the low-dimensional systems, and a convolutional network with $16$ channels for the spatio-temporal ones. For these three baselines, we sweep the diffusion coefficient $\sigma_0 \in \{0, 0.01, 0.05\}$ and keep the best value for each experiment and ablation.

For the Gray-Scott experiment, we adapted the interpolants, which are defined pointwise in time and since they do not have mechanisms to handle a pixel that is missing in the frames they interpolate between. We therefore fill the missing pixels of each observed frame with the mean of the observed pixels in a $3\times3$ neighbourhood, before the interpolation. The same fill is applied to all four interpolants, so it favours none of them.

\subsection{Hyperparameters} \label{app:hparams}
Table~\ref{tab:hparams} reports the main hyperparameters of our method for each system. All models are trained with AdamW, with a weight decay of $5\cdot10^{-7}$ for the low-dimensional systems and $10^{-6}$ for the spatio-temporal ones. The KL weight $\beta$ is annealed linearly from $0$ over the first $1000$ steps. For every trajectory, we draw $10$ query times per training step. The bounds of the log-uniform buffer gaps are set from the training data (\S\ref{sec:teacher}): $\delta_{\min}$ is the smallest inter-observation gap and $\delta_{\max}$ three times the median gap. For the spatio-temporal systems, we additionally clip the gradient norm to $1$ and use a free-bits threshold of $0.05$ on the KL term. We use early stopping on the validation error over the Early horizon; for ERA5, the model is trained for a fixed number of epochs.

\begin{table}[h]
\centering
\small
\caption{Main hyperparameters of our method. The baselines use the same velocity-field network, history length and optimisation settings. MLP: $4$ hidden layers of width $128$; CNN: residual convolutional network, with base channels / residual blocks.}
\label{tab:hparams}
\resizebox{\textwidth}{!}{%
\begin{tabular}{lcccccc}
\toprule
 & Lorenz & RLC & SWE & RD & Gray-Scott & ERA5 \\
\midrule
t-INR encoder & Transformer & Transformer & (2+1)-D CNN & (2+1)-D CNN & (2+1)-D CNN & (2+1)-D CNN \\
Velocity field & MLP & MLP & CNN $32/2$ & CNN $32/2$ & CNN $32/2$ & CNN $64/3$ \\
Latent dim.\ $m$ & $32$ & $8$ & $64$ & $64$ & $64$ & $128$ \\
History length $h$ & $10$ & $10$ & $10$ & $10$ & $10$ & $10$ \\
VF learning rate & $10^{-3}$ & $5\cdot10^{-4}$ & $10^{-4}$ & $10^{-4}$ & $10^{-4}$ & $10^{-4}$ \\
Encoder learning rate & $10^{-4}$ & $3\cdot10^{-5}$ & $10^{-5}$ & $10^{-5}$ & $10^{-5}$ & $10^{-5}$ \\
KL weight $\beta$ & $0.1$ & $0.1$ & $0.1$ & $0.1$ & $0.1$ & $0.1$ \\
Batch size & $25$ & $25$ & $10$ & $10$ & $10$ & $16$ \\
Max.\ epochs & $2000$ & $2000$ & $2000$ & $2000$ & $2000$ & $2000$ \\
ODE solver & RK4 & Dopri5 & RK4 & RK4 & RK4 & RK4 \\
\bottomrule
\end{tabular}}
\end{table}

\subsection{Compute resources} \label{app:compute}
Each model is trained on a single GPU. For the low-dimensional systems (Lorenz, RLC), training one model for one seed takes at most about $4$ hours on a GPU with $12$\,GB of memory. For the spatio-temporal systems and ERA5, training one model for one seed takes about $12$ hours on a GPU with at least $32$\,GB of memory, and up to about $50$\,GB for the largest configurations. These costs are approximate and apply to each seed.

\section{Ablation: how the history buffer is sampled} \label{sec:ablation_sampler}
\begin{table}[t]
\centering
\caption{Ablation on the history-buffer sampling strategy for the INR: log-uniform gaps before $\tau$ vs. sampling from the on-grid discrete buffer, across training history densities.}
\label{tab:table_on_grid}
\begin{tabular}{llccc}
\toprule
& & \multicolumn{3}{c}{Training observations density} \\
\cmidrule(lr){3-5}
Dataset & History sampling & $12.5\%$ $\downarrow$ & $25\%$ $\downarrow$ & $50\%$  $\downarrow$ \\
\midrule
\multirow{2}{*}{RLC}
& Log-uniform  & $\mathbf{0.52} \pm 0.07$ & $\mathbf{0.38} \pm 0.03$ & $\mathbf{0.14}\pm 0.01$ \\
& On-grid buffer     & $0.85 \pm 0.04$ & $1.03 \pm 0.02$ & $\mathbf{0.13} \pm 0.01$ \\
\midrule
\multirow{2}{*}{Lorenz}
& Log-uniform  & $\mathbf{0.31} \pm 0.09$ & $\mathbf{0.25} \pm 0.09$ & $\mathbf{0.29} \pm 0.14$ \\
& On-grid buffer     & $0.33 \pm 0.08$ & $0.30 \pm 0.08$ & $0.53 \pm 0.03$ \\
\midrule
\multirow{2}{*}{Shallow Water}
& Log-uniform  & $\mathbf{0.07} \pm 0.01$ & $\mathbf{0.02} \pm 0.01$ & $0.03 \pm \text{\small $\expnumber{1}{-3}$}$ \\
& On-grid buffer     & $0.08 \pm \text{\small $\expnumber{1}{-3}$}$ & $0.06 \pm \text{\small $\expnumber{1}{-3}$}$ & $\mathbf{0.03} \pm \text{\small $\expnumber{1}{-3}$}$ \\
\midrule
\multirow{2}{*}{Reaction-Diffusion}
& Log-uniform  & $\mathbf{0.13} \pm 0.03$ & $\mathbf{0.08} \pm \text{\small $\expnumber{1}{-3}$}$ & $0.13 \pm 0.03$ \\
& On-grid buffer     & $0.15 \pm \text{\small $\expnumber{1}{-3}$}$ & $0.10 \pm \text{\small $\expnumber{1}{-3}$}$ & $\mathbf{0.12} \pm 0.01$ \\
\bottomrule
\end{tabular}
\end{table}

Because the frozen teacher can be queried anywhere, we are free to choose where the history
buffer is placed during training, and the choice matters more than we expected. Sampling
buffer times with log-uniform gaps is compared in Table~\ref{tab:table_on_grid} against
taking the buffer on the observed grid. Log-uniform gaps expose the velocity
field to buffer points spanning several orders of magnitude, including close points and
widely spaced buffers, whereas an on-grid buffer is always
as dense as the training data and leaves the model unprepared for them. Notably, the
ordering can reverse at high density, where the on-grid buffer can be a better
choice.

\clearpage
\section{Forecast visualisations} \label{app:forecast_vis}
We show exmaples of test trajectories forecast at $25\%$ density by our method and by the strongest baseline of each spatio-temporal system in Table~\ref{tab:main}. Steps are counted from the start of the forecast.

\begin{figure}[h]
    \centering
    \includegraphics[width=\linewidth]{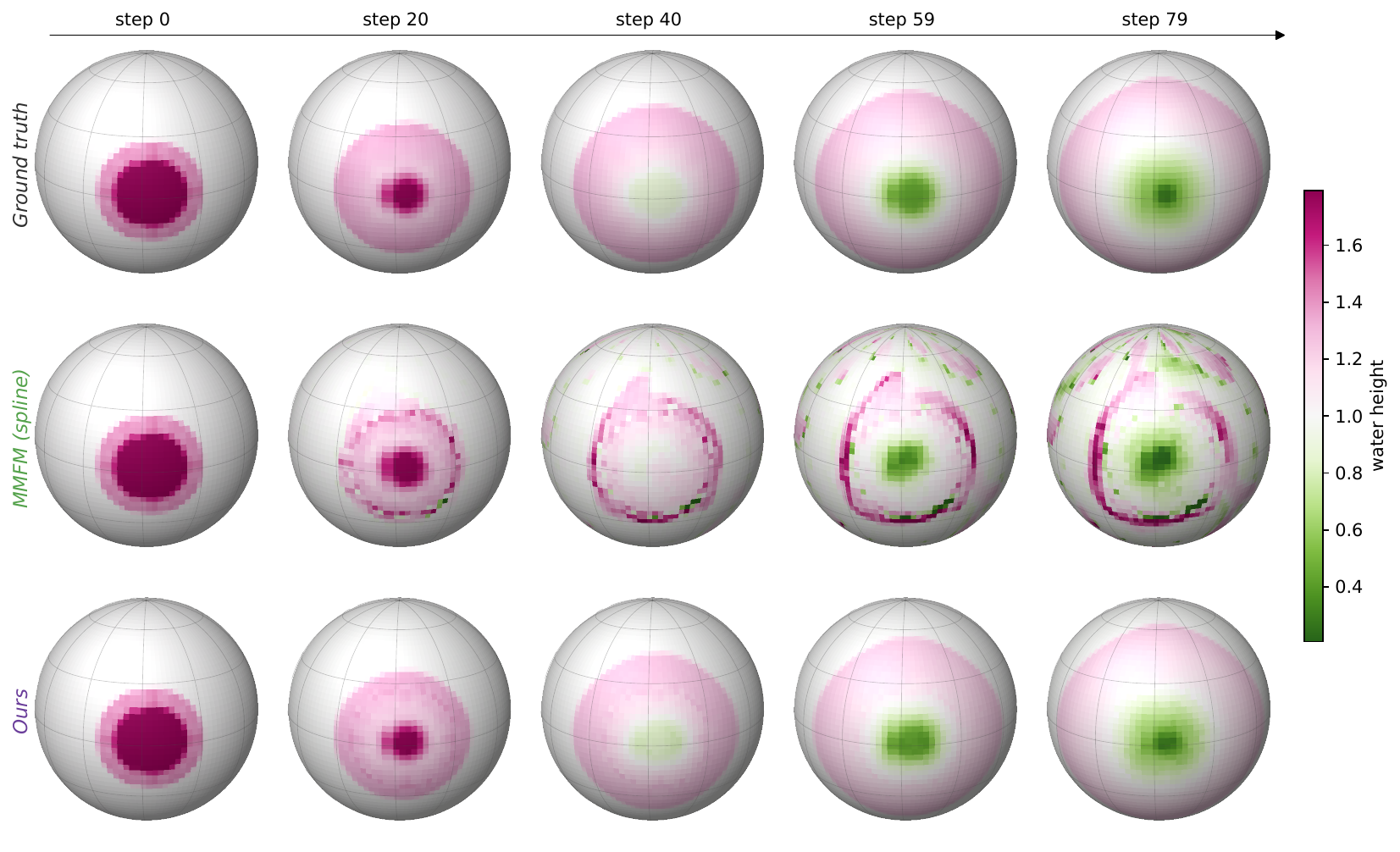}
    \caption{Shallow Water at $25\%$ density: ground truth, MMFM (the strongest baseline) and our method on a test trajectory. For readability the square domain is drawn on a sphere, and the colours show the water height around its rest level.}
    \label{fig:vis_swe}
\end{figure}

\begin{figure}[h]
    \centering
    \includegraphics[width=\linewidth]{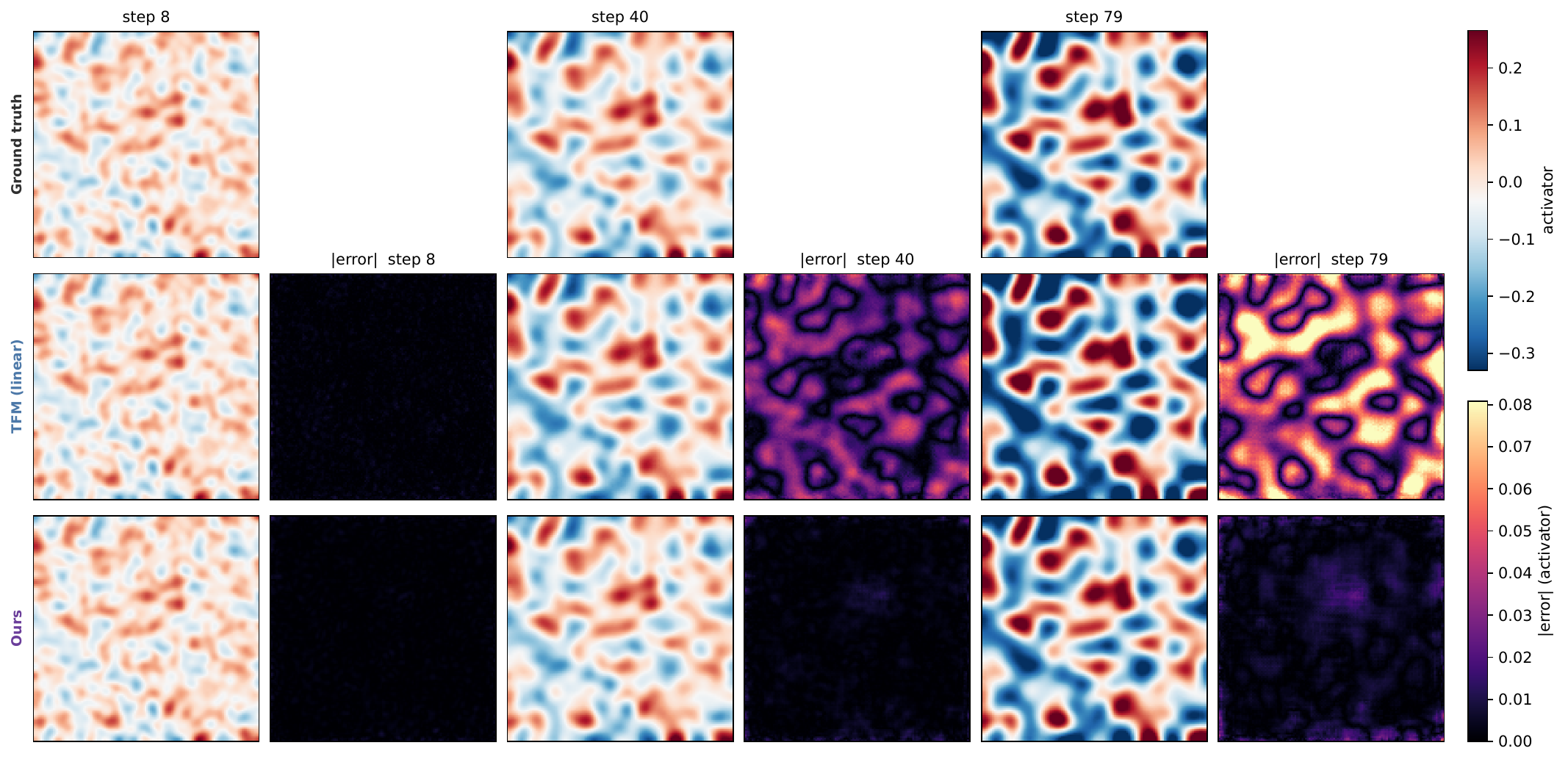}
    \caption{Reaction-Diffusion at $25\%$ density: activator field of one test trajectory at steps $8$, $40$ and $79$, for the ground truth, TFM (the strongest baseline) and our method, with the absolute error next to each forecast. The error of TFM grows over the whole domain as the pattern sharpens, whereas the error of our method remains close to zero over the horizon. Instead our method is able to capture those high frequency details.}
    \label{fig:vis_rd}
\end{figure}

\begin{figure}[h]
    \centering
    \includegraphics[width=0.75\linewidth]{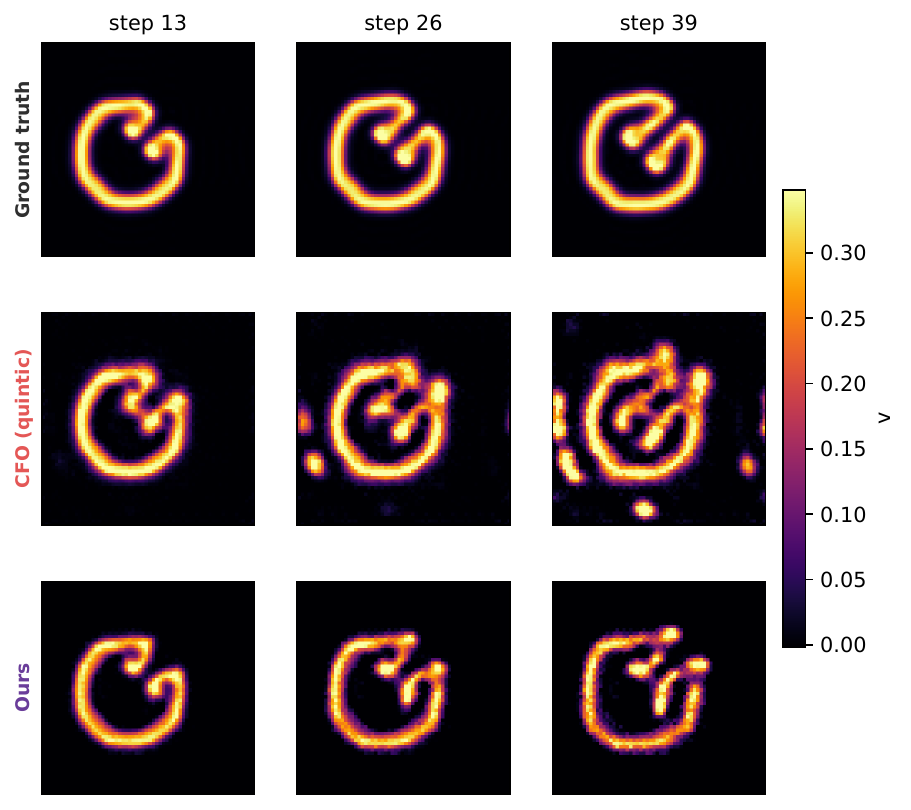}
    \caption{Gray-Scott at $25\%$ density in time and space: concentration $v$ of one test trajectory at steps different steps, for the ground truth, CFO (the strongest baseline) and our method.}
    \label{fig:vis_gs}
\end{figure}

\clearpage
\section{Weather forecasting on ERA5}
\label{app:era5}
\paragraph{Dataset.} We use ERA5 reanalysis on the WeatherBench \citep{weatherbench2020} $5.625^\circ$ grid ($32\times64$), with five variables: geopotential at 500\,hPa (z500), temperature at 850\,hPa (t850), 2\,m temperature (t2m), and the two 10\,m wind components (u10, v10). The data are taken at a $6$\,h cadence and cut into non-overlapping windows of $15$ days ($60$ frames). The years 1979--2015 are used for training, 2016 for validation and 2017--2018 for test, which gives $900$, $24$ and $48$ windows. In each training window we keep $50\%$ of the frames ($30$ of $60$, mean gap $12$\,h) at irregular times. Fields are standardised per channel, and all metrics are reported back in physical units.

\paragraph{Models.} The t-INR follows the spatio-temporal architecture of App. ~\ref{app:tinr}, with $5$ input channels. The velocity field is the convolutional network of App. ~\ref{app:vf}, enlarged to $64$ base channels and $3$ residual blocks, with a latent of dimension $128$ (Table~\ref{tab:hparams}). The baselines use the same velocity-field network. All models are trained for $2000$ epochs over three seeds.

\paragraph{Evaluation.} From a history of $h=10$ observations, each model produces a single rollout over the remaining $40$ frames of the window, from $0$ to $234$\,h lead, scored every $6$\,h against the reanalysis. We report the $\cos(\mathrm{lat})$-weighted RMSE, the Anomaly Correlation Coefficient (ACC), the CRPS and the nMSE. The climatology is computed, for each variable and grid point, as the mean of the field over all the time steps of the training years (1979--2015). It serves both as a reference forecast, which predicts this mean at every lead time, and as the reference for the ACC, which is computed on the anomalies with respect to it. The SDE baselines are scored on the mean of an ensemble of $20$ forecasts, and our method on the forecast obtained with the posterior mean of $\mathbf{z}$. The CRPS uses the same ensembles of $20$ members; for our method, the spread of the latent ensemble is calibrated on the validation year. As references, we report persistence, which carries the last observed state forward, and climatology.

\begin{figure}[h]
    \centering
    \includegraphics[width=\linewidth]{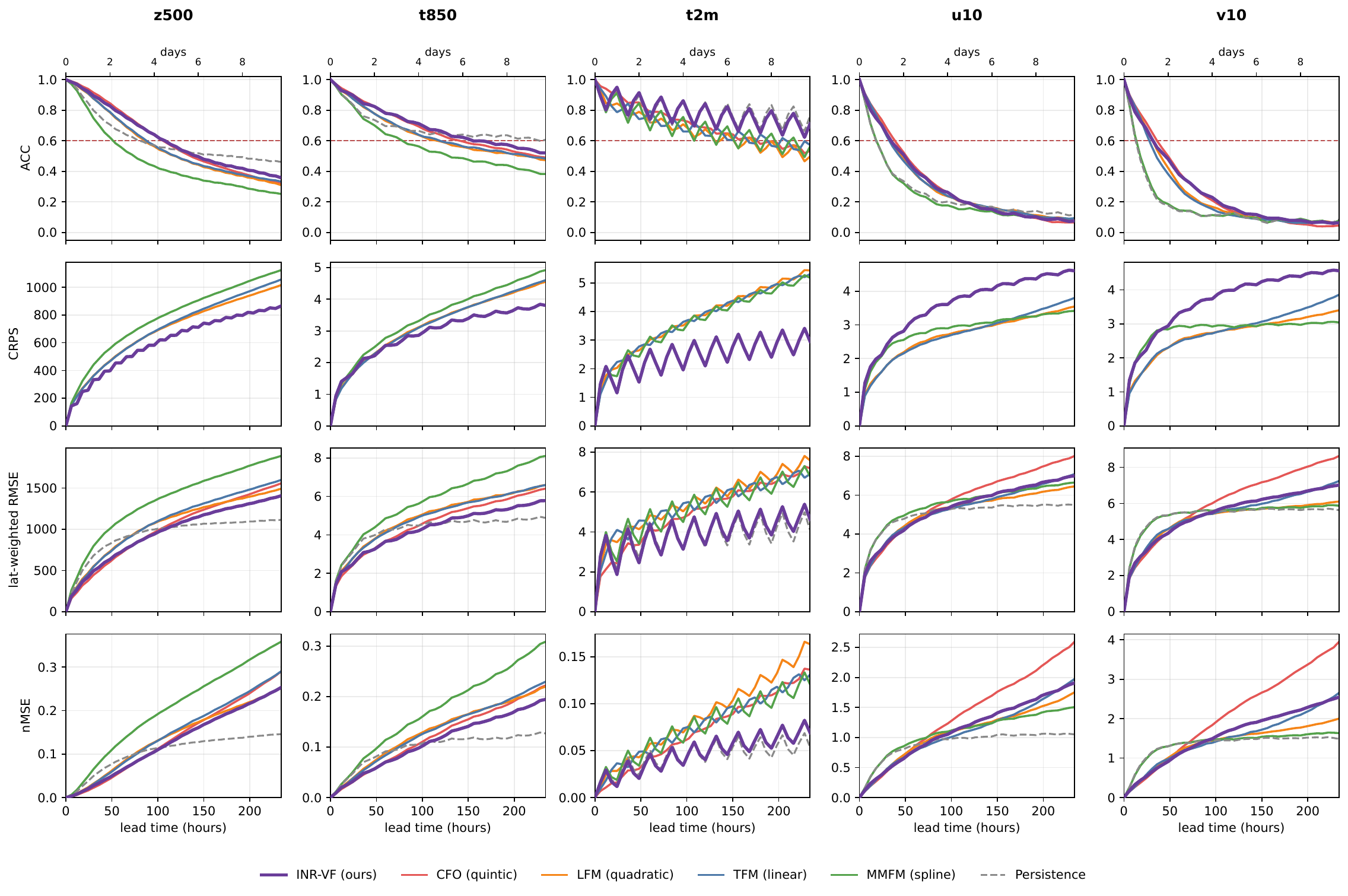}
    \caption{ERA5 skill against lead time, until $234$\,h.}
    \label{fig:era5_skill_full}
\end{figure}

\begin{figure}[b]
    \centering
    \includegraphics[width=\linewidth]{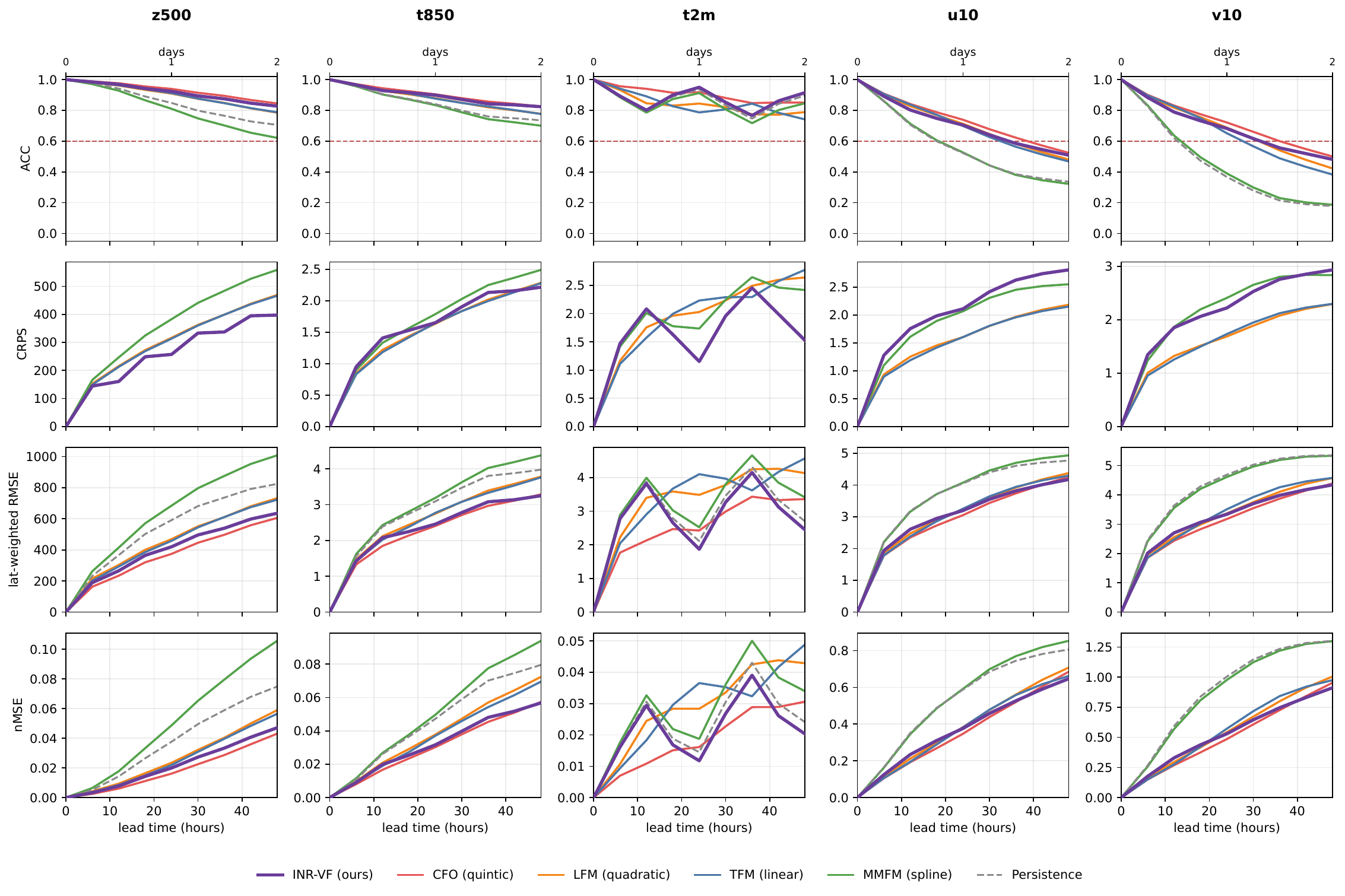}
    \caption{ERA5 skill against lead time, until $48$\,h.}
    \label{fig:era5_skill_48}
\end{figure}

\begin{figure}[b]
    \centering
    \includegraphics[width=\linewidth, height=7cm]{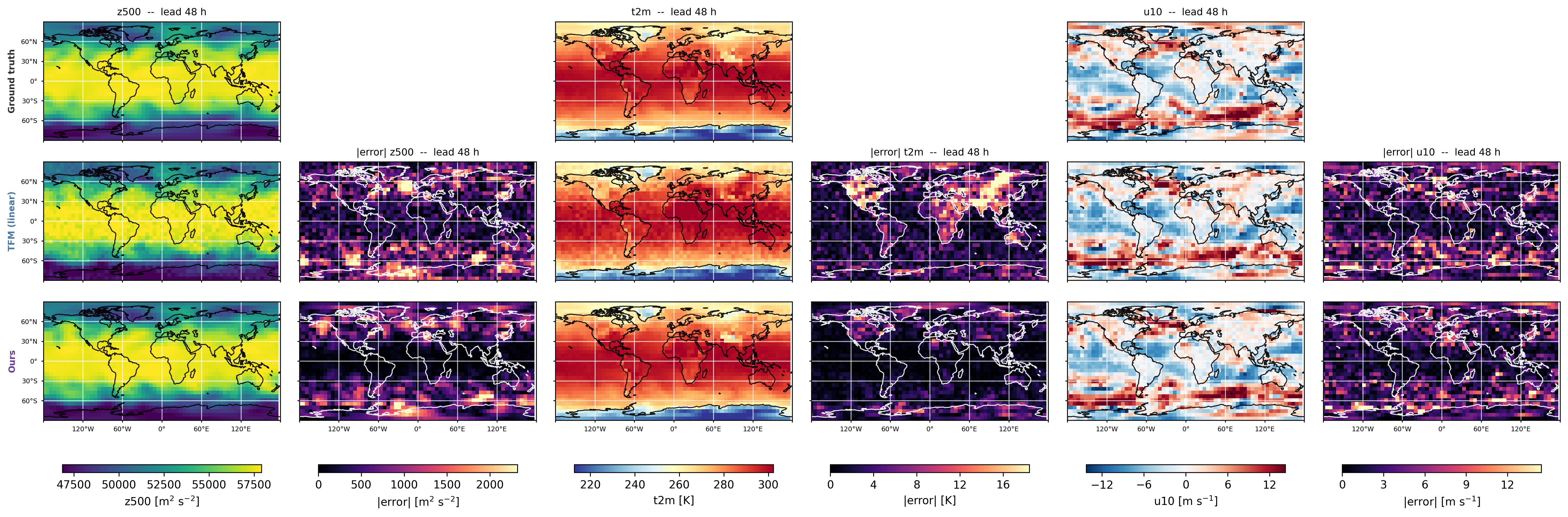}
    \caption{ERA5 ground truth test sample and forecasts from TFM and our method at a lead time of $48$\,h, with the corresponding prediction errors.}
    \label{fig:era5_skill_48_vis}
\end{figure}

\end{document}